\documentclass{article}
\usepackage{rotating}
\usepackage{kotex}
\usepackage{PRIMEarxiv}

\usepackage[utf8]{inputenc} 
\usepackage[T1]{fontenc}    
\usepackage{hyperref}       
\usepackage{url}            
\usepackage{booktabs}       
\usepackage{amsfonts}       
\usepackage{nicefrac}       
\usepackage{microtype}      
\usepackage{multirow}       
\usepackage{float}          
\usepackage{lipsum}
\usepackage{amsmath} 
\usepackage{fancyhdr}       
\usepackage{graphicx}       
\graphicspath{{media/}}     

\title{The selective use of physics knowledge in policy: how interdisciplinary physics bridges subfields and shapes policy influence}

\author{
  Jeongmin Lee\textsuperscript{1}, Jisung Yoon\textsuperscript{1*} \\\\
  KDI School of Public Policy and Management, Sejong-si, Republic of Korea \\
  \\
  \texttt{Corresponding author. E-mail: *jsyoon@kdischool.ac.kr} \\
}

\begin{document}
\maketitle

\begin{abstract}
Scientific knowledge has become increasingly central to policymaking as societies confront challenges related to technological change, climate risk, and public health. Despite the growing emphasis on evidence-based policy, a systematic understanding of how science is selectively used in policy—specifically which forms of knowledge are preferred and which scientific citations translate into influence, remains limited. We address these questions by constructing a novel dataset that links policy documents from the Overton database with publications from the American Physical Society, enabling a fine-grained analysis of how physics knowledge enters and circulates in policy discourse. Using detailed subfield classifications, we provide quantitative evidence for a persistent gap between scientific communities and policymakers. First, we find that policy documents disproportionately draw on broad and \textit{interdisciplinary areas of physics}, such as General Physics and Interdisciplinary Physics, rather than mirroring the structure of physics research production. Second, we identify substantial institutional heterogeneity with systematic differences in subfield preferences across policy-producing organizations and topics. Third, network analysis reveals that interdisciplinary areas of physics act as a central bridge connecting specialized subfields. Finally, regression analysis reveals a clear separation between policy visibility and policy influence. While interdisciplinary areas of physics facilitate entry into policy discourse, it does not necessarily increase downstream policy influence. Conversely, documents citing geophysics are associated with approximately 24 percent higher policy influence, likely driven by the political salience of climate change policy. Our findings underscore the distinction between scientific visibility and policy influence, contributing to a deeper understanding of the complex relationship between scientific communities and policy systems.
\end{abstract}

\section{Introduction}
The question of how science is cited in policy and which forms of scientific knowledge shape policy outcomes has become increasingly critical, yet systematic empirical evidence remains limited \cite{jensen2025can}. As societies confront a convergence of existential challenges—ranging from the regulation of artificial intelligence and the mitigation of climate change to the management of global pandemics—, the demand for rigorous evidence-based policymaking has intensified \cite{Yin2021,bornmann2022relevant,Wilson,Cabral,asatani2025influential}. This surge is not merely quantitative; it reflects a fundamental shift in governance where scientific authority is increasingly mobilized to legitimize decisions, navigate uncertainty, and design complex regulatory frameworks\cite{dorta2024societal,cairney2016politics, Weiss}. This shift is reflected in the growing prevalence of academic references in policy documents, which rose from under 20 percent in the mid-1990s to more than 35 percent by 2020 \cite{Fang2024}. However, citations from policy to academia serve as more than just a simple indicator of knowledge diffusion; they reflect complex, strategic engagement patterns where policy documents utilize direct references and indirect citations to construct validity and bridge epistemic gaps \cite{asatani2025influential, Cao, WAGNER201114}.

Despite the growth in citation volume, the \textit{two communities} theory highlights a persistent and structural gap between scientists and policymakers \cite{caplan1979two}. Scientific research is largely supply-driven, operating on long time horizons and organized around internal disciplinary priorities, whereas policymakers are demand-driven, operating under tight political constraints and focusing on socially defined problems that require immediate, actionable solutions \cite{schoor2021science, ruhl2019engaging}. The resulting structural divergence means that scientific evidence is rarely incorporated as a neutral or comprehensive input into the decision-making process. Instead, scientific findings are selectively mobilized and interpreted to align with institutional mandates, reinforce political legitimacy, or substantiate pre-existing policy preferences \cite{Head,John2023,cairney2016politics,Brian,hahn2019building}. Consequently, the mobilization of scientific advice is highly segmented: government agencies tend to emphasize practical environmental and regulatory research to support implementation \cite{yin2022public, bozeman2000technology, an2022contingent}, intergovernmental organizations (IGOs) prioritize broad consensus on global issues like security \cite{lopes2025effects}, and think tanks frequently draw on specialized or ideological research to advocate for specific network-based policy agendas \cite{allern2020role,teitz2009analysis,Furnas2025}.

Nonetheless, the mechanisms of selective mobilization of scientific knowledge remain obscured~\cite{yin2022public}. 
While policy demand appears to broadly mirror scientific supply in fields like climate science~\cite{bornmann2022relevant}, it is unclear whether such correspondence persists in other disciplines where societal relevance is mediated rather than direct. In this context, physics subdisciplines serve as a strategic hard test for the integration of scientific knowledge and policy advice; unlike medicine or economics, the impact of physics is often filtered through technology, resulting in significantly lower citation rates compared to other disciplines~\cite{Wilson, asatani2025influential, fanelli2013bibliometric, salter2001economic}. Focusing on a discipline structurally distant from immediate regulatory needs isolates the bridging mechanisms required to span the science-policy divide. This approach shifts the inquiry from \textit{whether} science is mobilized to \textit{how} abstract knowledge is selectively translated for policy discourse.

Given the structural distance between the discipline of physics and the policy sphere, \textit{interdisciplinarity} emerges as a critical mechanism for knowledge translation. Within physics, interdisciplinary areas play a central role by structurally linking specialized subfields, effectively acting as the cohesive glue of the network~\cite{sinatra2015century}. Similarly, as policymakers confront increasingly complex societal challenges, interdisciplinary research—by integrating methods and concepts from multiple fields—is often perceived as more actionable and relevant than highly specialized work~\cite{Porter2009,shaman2013fostering,Hu2024,KWON2022121767,DESTE2019103799,Atypical}. We hypothesize that such interdisciplinary research acts as a crucial gateway, rendering abstract technical knowledge legible to policy audiences by linking fundamental concepts to broader societal concerns~\cite{DESTE2019103799,Pan2012,sinatra2015century,Lutz}. However, while this gateway function lowers entry barriers, there is a critical distinction between \textit{visibility} and \textit{influence}. While interdisciplinarity may facilitate the incorporation of physics knowledge into policy discourse (\textit{visibility}, measured by citations from policy documents to academic research), it remains unclear whether such integration translates into tangible downstream policy impact (\textit{influence}, measured by the number of citations received by policy documents that cite academic research)~\cite{asatani2025influential,LEMOS200557,Pflanzer2023}.

To empirically test our hypothesis and disentangle visibility from influence, we construct a linked dataset matching American Physical Society (APS) publications to policy documents in the Overton database. First, we identify a pronounced structural mismatch: policy actors disproportionately cite \textit{interdisciplinary areas of physics}, whereas the specialized subfields dominating scientific output receive comparatively little attention. Second, our observed divergence confirms that policy actors operate under a demand for evidence that is broadly applicable and easily interpretable, favoring interdisciplinary bridges over specialized cores \cite{Briggs}. Third, network analysis reveals that interdisciplinary areas of physics serve as structural intermediates that connect specialized disciplines. Finally, regression analysis shows that while policy documents frequently cite interdisciplinary areas of physics—demonstrating high visibility—reports citing interdisciplinary areas of physics do not exhibit significantly greater downstream influence.

\section{Materials and Methods}
\subsection{Data}
We construct our dataset by linking policy documents from the Overton database to scientific publications from the American Physical Society (APS). On the policy side, we use Overton, a large-scale curated repository that systematically tracks citation links from policy texts—such as reports, guidelines, and white papers—to academic publications. It encompasses documents written for or by policymakers across governments, intergovernmental organizations (IGOs), and think tanks \cite{overton_definition}. Compared to alternative sources such as Altmetric, Overton provides a broader and more systematic coverage of policy-science linkages, enabling a more reliable assessment of how scientific knowledge is incorporated into policy discourse \cite{szomszor2022overton}. To ensure consistency, we restrict our analysis to policy documents produced by United States entities (government, legislative bodies, and think tanks) and international IGOs.

Beyond citation linkages, Overton provides rich bibliographic metadata, including producing institutions, policy topics, and document summaries generated by large language models, spanning the period from 1995 to 2025. Importantly, it records citations between policy documents, for studying how to situate scientific references within the broader structure of policy influence. While citation motivations vary—ranging from evidentiary support to supplementary referencing—we focus on aggregate patterns to characterize macro-level trends rather than interpreting individual document-level contexts \cite{yu2023can}.

On the science side, we utilize publication data from the American Physical Society (APS), covering the period from 1978 to 2015. We classify the scientific content of these papers using the Physics and Astronomy Classification Scheme (PACS). Although APS replaced PACS with Physics Subject Headings (PhySH) in 2016, we retain PACS for this analysis to ensure historical consistency throughout our study period~\cite{sinatra2015century}. Unlike PhySH, which functions as a flexible concept graph, PACS provided the standard author-assigned taxonomic structure for physics research throughout the decades covered by our dataset. Essentially, its fixed hierarchical nature enables the systematic aggregation of granular research topics into stable and broad subfields~\cite{sinatra2015century, Pan2012}. We classify the papers into ten major categories based on the first digit of their PACS code:

\begin{itemize}
    \item Category 0 (GP): General Physics
    \item Category 1 (EPF): Elementary Particles and Fields
    \item Category 2 (NP): Nuclear Physics
    \item Category 3 (AM): Atomic and Molecular Physics
    \item Category 4 (EOA/HCF): Electromagnetism, Optics, Acoustics, Heat Transfer, Classical Mechanics, and Fluid Dynamics
    \item Category 5 (GP\&E): Gases, Plasmas, and Electric Discharges
    \item Category 6 (CM-SMT): Condensed Matter: Structural, Mechanical, and Thermal
    \item Category 7 (CM-EMO): Condensed Matter: Electronic, Electrical, Magnetic, and Optical
    \item Category 8 (IPR): Interdisciplinary Physics and Related Areas of Science and Technology
    \item Category 9 (GAA): Geophysics, Astronomy, and Astrophysics
\end{itemize}

In particular, we define Category 0: General Physics and Category 8: Interdisciplinary Physics as the primary domains of \textit{interdisciplinary areas of physics} within our framework~\cite{sinatra2015century, Pan2012}. We adopt this terminology to distinguish the collective functional category from the specific title of Category 8. Specifically, category 0 encompasses statistical physics and complex systems theory (subfield 05)—the mathematical foundation for analyzing social networks and systemic risk. Complementing this, Category 8 explicitly covers intersections with other scientific domains, including biological physics (87), renewable energy materials (81, 82), and the physics of social and economic systems (89). We hypothesize that these interdisciplinary areas of physics function as structural ``gateways,'' mediating the transfer of abstract physical science into the policy sphere.

\subsection{Structure of policy–science linkages}

\begin{figure}
  \centering
  \includegraphics[width=0.9\linewidth]{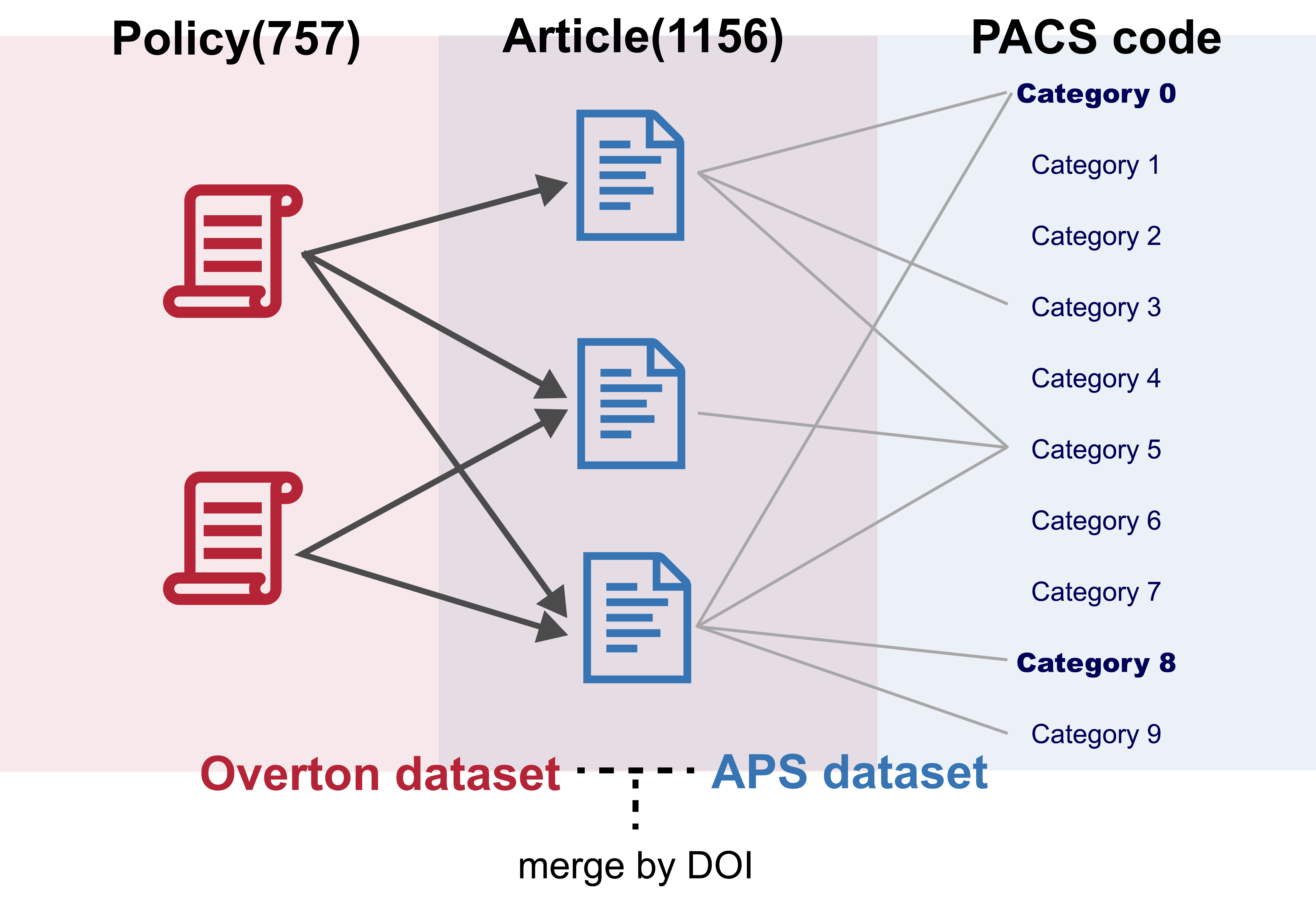}
    \caption{\textbf{Schematic of the data linkage and classification pipeline.} Policy documents from the Overton database (left) are linked to scientific articles in the American Physical Society corpus (center) via Digital Object Identifiers (DOIs). This linkage captures the structure of policy demand, where a single policy document may reference multiple scientific papers. Each cited paper is then characterized by its scientific content (right) using the first-level categories of the Physics and Astronomy Classification Scheme (PACS 0–9). Interdisciplinary areas of physics—specifically Category 0: General Physics and Category 8 :Interdisciplinary Physics and Related Areas of Science and Technology—are highlighted in bold.}
  \label{fig:schematic}
\end{figure}

Figure \ref{fig:schematic} illustrates the construction of our linked dataset, which systematically connects policy demand to scientific supply. We identify 757 policy documents from the Overton database containing citations to academic literature and link them via Digital Object Identifiers (DOIs) to 1,156 unique physics articles in the American Physical Society (APS) corpus. Our data capture the directional flow of information, where a single policy document frequently cites multiple scientific articles to substantiate their arguments. We characterize this one-to-many relationship as the policy demand for specific forms of scientific evidence.

The right panel illustrates how we characterize the scientific content of these citations using the Physics and Astronomy Classification Scheme (PACS). Each APS article is assigned up to five PACS codes \cite{enduri2022empirical,radicchi2011rescaling}, which map the research to specific subfields ranging from Category 0 to Category 9. For instance, a policy document addressing renewable energy might cite a paper on solar cell efficiency; this citation would logically link to relevant subfields such as Category 8: Interdisciplinary Physics or Category 7: Condensed Matter. This hierarchical structure provides a quantitative measure not just of the aggregate volume of citations but also the specific disciplinary composition of physics knowledge mobilized within policy discourse.

\subsection{Quantifying subfield representation in policy documents}

As policy documents often cite multiple papers and individual papers may be associated with multiple PACS codes, raw counts can bias the estimation of subfield prominence (see~Fig.S3). To mitigate biases arising from varying citation practices, we employ a fractional weighting scheme. Raw citation counts can misleadingly inflate the prominence of subfields if they rely heavily on policy documents with extensive reference lists or on multidisciplinary papers with numerous classification codes.

We therefore normalize the contribution of each citation at two levels: the policy document level and the individual paper level. Formally, let a policy document cite $n$ distinct physics papers. Each cited paper is assigned a base weight of $1/n$ to ensure that every policy document contributes equally to the aggregate analysis, regardless of its bibliography length. Next, if a specific cited paper is associated with $k$ PACS codes, its weight is evenly distributed among them, allocating $1/k$ to each category. The final contribution of a cited paper to a specific PACS category is the product of these weights:$$w = \frac{1}{n} \times \frac{1}{k}.$$For instance, consider a policy document that cites two physics papers ($n=2$). Paper $a$ is classified under four PACS codes (Categories 2, 3, 4, and 8), meaning $k_a=4$. Paper $b$ is classified under two PACS codes (Categories 4 and 8), meaning $k_b=2$. The total contribution of this policy document to Category 8 is calculated as the sum of the fractional weights from both papers:$$\text{Contribution} = \left( \frac{1}{2} \times \frac{1}{4} \right) + \left( \frac{1}{2} \times \frac{1}{2} \right) = 0.125 + 0.25 = 0.375,$$ ensuring the measured prominence of a subfield reflects specific and concentrated demand rather than incidental accumulation of broad or voluminous citations.

\subsection{Characterizing policy themes via topic modeling}

To systematically characterize the thematic context in which physics research is cited, we employ Latent Dirichlet Allocation (LDA), a generative probabilistic model that represents documents as mixtures of latent topics \cite{LDA}. LDA topic modeling identifies dominant semantic themes—such as healthcare, energy, or defense—without imposing predefined categories. We construct the textual corpus using metadata from the Overton database, concatenating the title and the LLM-generated summary for each policy document. After preprocessing the text—which includes lowercasing, tokenization, and filtering stop words and filtering non-alphanumeric characters—we utilize a Term Frequency–Inverse Document Frequency (TF-IDF) weighting scheme to mitigate the influence of high-frequency terms and extract discriminative topics~\cite{salton1975vector}. We set the number of topics—the LDA's hyperparameter—by balancing quantitative metrics of topic coherence and perplexity against qualitative interpretability. We identify six topics that best capture distinct policy domains (see~Fig.S4). We further validate our configuration using a Principal Component Analysis (PCA) projection, which demonstrates a clear separation of the identified themes within the latent semantic space (see~Fig.S6).

\subsection{Backbone extraction}
We construct a weighted co-citation network of PACS codes in policy documents to characterize the structural relationships among physics subfields. However, as the original network is dense and noisy, we extract the structural backbone using the disparity filter~\cite{Backbone}. The disparity filter evaluates the statistical significance of each link against a null hypothesis where a node's total weight is uniformly distributed across its connections. For a node $i$ with degree $k_i$ and total strength $s_i = \sum_{j} w_{ij}$, the normalized weight of the link to node $j$ is $p_{ij} = w_{ij}/s_i$. Then, statistical significance of each link $\alpha_{ij}$ is:$$\alpha_{ij} = 1 - (k_{i} - 1) \int_{0}^{p_{ij}} (1 - x)^{k_{i} - 2} \,dx.$$ We retain only those links that meet a significance threshold of $\alpha_{ij} < 0.1$. All subsequent topological analyzes are performed on the resulting backbone network, ensuring that our results reflect the most non-trivial and robust associations between subfields.

\subsection{Community detection}
To uncover the meso-scale functional structure between physics subdisciplines, we apply the Louvain algorithm~\cite{Louvain} to the co-citation network of PACS codes constructed from policy documents. The Louvain algorithm partitions the graph by maximizing modularity $Q$, a metric that quantifies the density of weighted edges within communities relative to a null model that preserves node strengths. We define modularity formally as:$$Q = \frac{1}{2m} \sum_{i,j} \left( A_{ij} - \frac{k_i k_j}{2m} \right) \delta(c_i, c_j)$$Here, $A_{ij}$ represents the edge weight between nodes $i$ and $j$, $k_i$ denotes the weighted degree (strength) of node $i$, $m$ is the total network weight, and $\delta(c_i, c_j)$ is the Kronecker delta, equal to 1 if nodes $i$ and $j$ share the same community membership and 0 otherwise. With the standard resolution parameter ($\gamma = 1$), we find five communities with modularity of $Q = 0.46$, indicating a robust community structure with well-defined clusters of co-cited subfields.

\subsection{Regression analysis: visibility vs. influence}
To examine the \textit{visibility} and \textit{influence} of the physics subdisciplines in policy contexts, we introduce two distinct metrics characterizing the science-policy interface. We define \textit{visibility} as the number of citations a physics subdiscipline receives from policy documents, representing the direct uptake or presence of scientific knowledge within the policy sphere. In contrast, we define \textit{influence} as the downstream impact of these policy documents, quantified by the number of citations received by the policy reports that cite specific academic research. To rigorously test both concepts, we introduce two Ordinary Least Squares (OLS) regression models as follows.

In the first model, we estimate the \textit{visibility} of each physics subdiscipline rigorously. To this end, we define the dependent variable $\log(\text{PaperCitations}_i)$, as the count of citations an academic paper $i$ receives from policy documents. Given the heavy-tailed distribution of citations, we apply a natural logarithmic transformation (see~Fig.S3). The primary independent variables are binary indicators $\text{Category}_{ik}$, which denote whether the paper $i$ belongs to the physics subfield $k$ based on first-level PACS codes. We also include additional independent variables that capture the core scientific characteristics, specifically the disruption score of the paper ($\text{Disruption}_i$) ~\cite{funk2017dynamic, sciscinet} as a measure of novelty and $\log(\text{AcademicCitations}_i)$ as a proxy of established scientific influence ~\cite{wang2013quantifying}. Furthermore, we incorporate production-related independent variables, including an indicator for top-author reputation ($\text{TopAuthor}_i$)—which identifies whether a top 1\% author is included in the author list (see~Table.S12)—and team size ($\text{NumAuthors}_i$). Finally, we include a vector of control variables, $\mathbf{C}_i$, comprising publication year and journal fixed effects to absorb temporal trends and venue-specific prestige (see~Fig.S1). The full specification is defined as:

\begin{align*}
\log(\text{PaperCitations}_i) =\;&
\alpha
+ \sum_{k=0}^{9} \beta_k \,\text{Category}_{ik}
+ \lambda_1 \,\text{Disruption}_i
\\&+ \lambda_2 \,\log(\text{AcademicCitations}_i)
+ \lambda_3 \,\text{TopAuthor}_i
\\&+ \lambda_4 \,\text{NumAuthors}_i
+ \mathbf{C}_i \boldsymbol{\delta}
+ \varepsilon_i .
\end{align*}
\\
In the second model, we examine the \textit{influence} of policy documents. The dependent variable, $\log(\text{PolicyCitations}_i)$ is the number of citations policy document $i$ receives from other policy documents, serving as a proxy for its \textit{influence within the policy sphere}. We also apply a natural transformation to account for the heavy-tailed distribution of citations (see~Fig.S3). Similarly to the first model, the primary independent variables are binary indicators $\text{Category}_{ik}$, which take the value of 1 if document $i$ cites at least one paper belonging to the PACS category $k$ (where $k = 0, \dots, 9$). To test the effect of the intensity of scientific use, we include $\text{NumPapers}_i$, the count of distinct APS papers cited by the document. We include a set of control variables, $\mathbf{C}_i$, to account for confounding factors. These control variables include fixed effects for publication year, institutional type (e.g., IGO vs. Think Tank), and the dominant policy topic derived from our LDA model. Additionally, we control for the scientific characteristics of the cited research papers, specifically the average academic citation count, the presence of the top 1\% elite authors and the average disruption score. 

\begin{align*}
\log(\text{PolicyCitations}_i) =\;&
\alpha
+ \sum_{k=0}^{9} \beta_k \,\text{Category}_{ik}
\\&+ \gamma \,\text{NumPapers}_i \nonumber 
+ \mathbf{C}_i \boldsymbol{\delta}
+ \varepsilon_i .
\end{align*}

\section{Results}

\subsection{Structural Divergence Between Scientific Supply and Policy Demand}

\begin{figure}[h]
  \centering
  \includegraphics[width=0.99\linewidth]{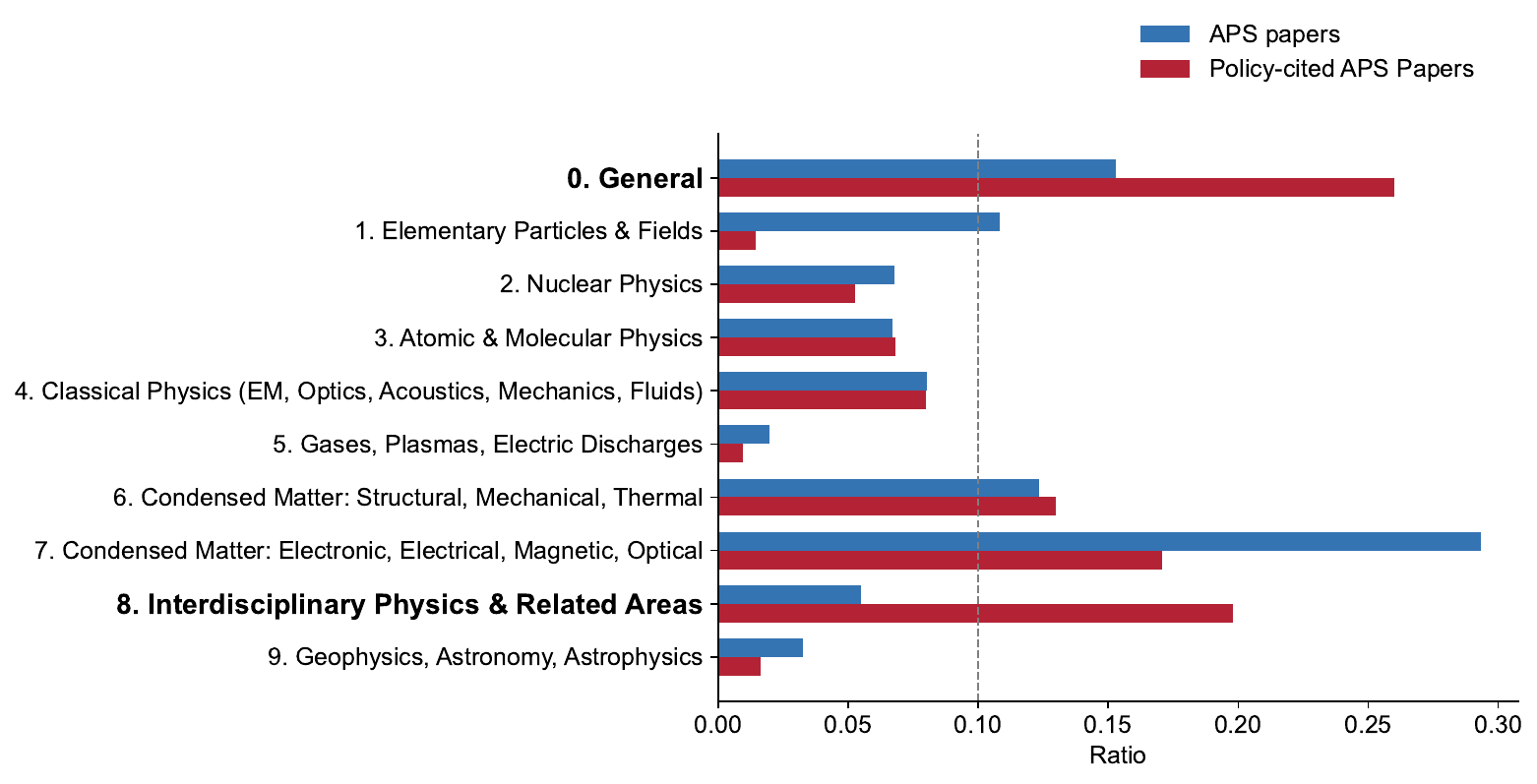}
  \caption{\textbf{Distribution of scientific supply versus policy demand.} The blue bars represent the full APS corpus (scientific supply), while the red bars denote the subset of papers cited in policy documents (policy demand). Frequencies are fractionally weighted to account for papers assigned to multiple PACS categories. The vertical dotted line marks the expected proportion under a uniform null model, representing the baseline where all ten categories are equally represented. The label for \textit{Interdisciplinary areas of physics} is highlighted in bold. While Category 7: Condensed Matter Physics dominates overall academic production, policy documents exhibit a substantial preference for \textit{Interdisciplinary areas of physics}, indicating a structural divergence between scientific supply and policy demand.}
  \label{fig:distribution}
\end{figure}

To quantify the alignment between scientific supply and policy demand, we compare the fractionally weighted distribution of PACS categories in the full APS publication corpus (1977–2015, $N=435,253$) against the subset of papers cited by policy documents. The academic landscape, represented by the blue bars in Figure \ref{fig:distribution}, reflects the internal priorities of the discipline. Scientific supply is heavily dominated by Category 7: Condensed Matter: Electronic, which accounts for 29.3\% of the total research output, and Category 1: Elementary Particles and Fields at 10.8\%. Together with Category 6: Condensed Matter: Structural (12.3\%), these fields form the traditional backbone of modern physics, prioritizing fundamental inquiries into the nature of matter and the universe's elementary constituents.

In contrast, the structure of policy demand—represented by the red bars—reveals a completely different set of priorities. Policy documents do not simply sample randomly from the available literature; instead, they selectively filter for broader and more integrative fields. The policy demand portfolio is dominated by interdisciplinary areas of physics, Category 0: General Physics and Category 8: Interdisciplinary Physics, which account for 26.0\% and 19.8\% of the policy-cited weight, respectively. The shift is most dramatic in Category 8, which represents a niche sector of academic production (5.5\%) but becomes a primary pillar of policy engagement. In contrast, Category 1: Elementary Particles, despite its high academic volume, is marginalized in the policy sphere, accounting for only 1.4\% of citations. 

Our comparison highlights a fundamental structural divergence: the supply-side hierarchy of physics does not map linearly onto the demand-side utility. Policy documents exhibit a strong preference for cross-cutting and methodologically adaptable frameworks over highly specialized, fundamental domains. While Condensed Matter Physics reigns supreme in academic output, its relative influence diminishes in the policy context. Instead, the focus shifts toward interdisciplinary areas of physics (Categories 0 and 8), suggesting that policymakers value physics most when it is packaged in frameworks that bridge disciplinary boundaries or apply directly to systemic complexity.

\subsection{Institutional and Thematic Heterogeneity in Policy Demand}

Does the aggregate mismatch between scientific supply and policy demand imply a uniform preference for generalist physics across the entire policy spectrum? Or does the demand for specific subfields vary depending on the problem being addressed and the organization framing the solution? We identify substantial heterogeneity, revealing that the ``mismatch'' is not a monolithic phenomenon but is structured by systematic differences in subfield preferences across policy-producing organizations and topics.

\begin{figure}[h]
    \centering
\includegraphics[width=0.8\linewidth]{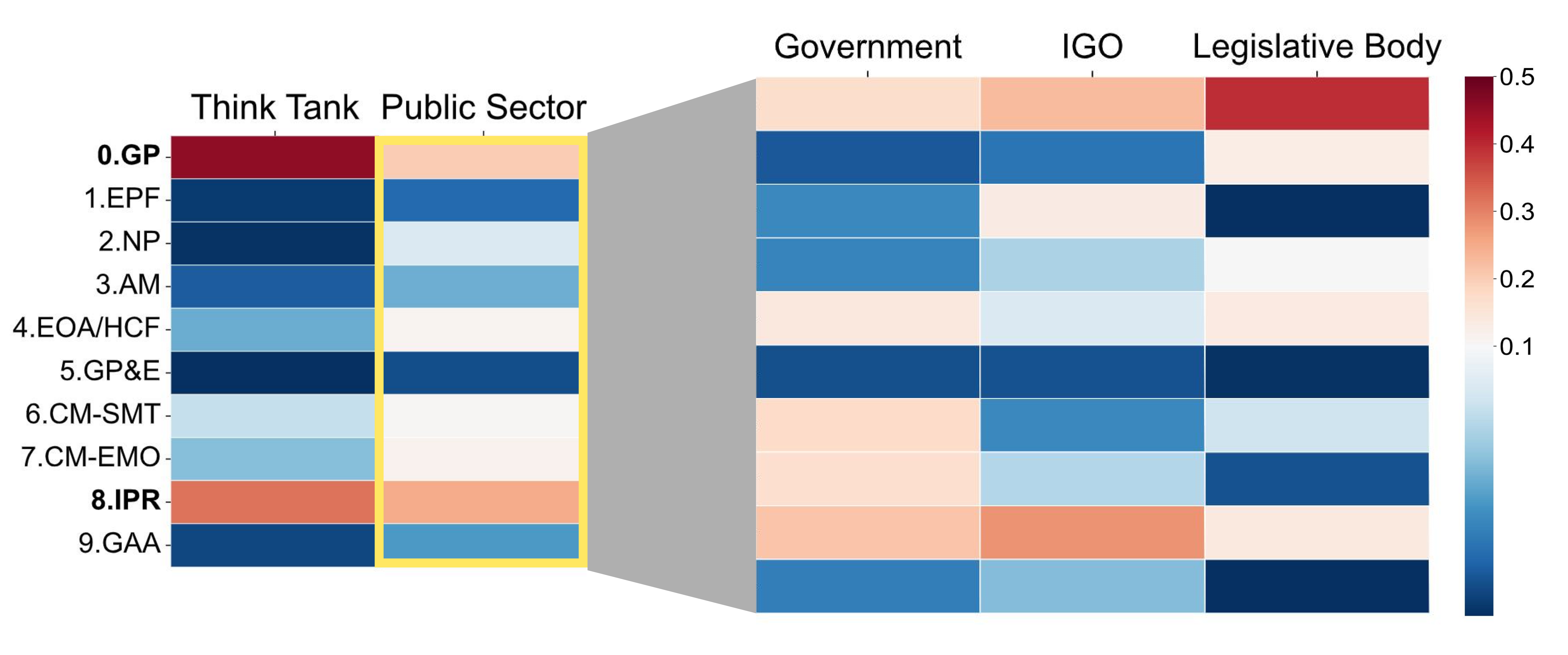}
    \caption{\textbf{Institutional heterogeneity in the use of physics subfields.} The heatmap displays citation intensity normalized by the number of cited papers and assigned PACS codes per policy document (See Materials and Methods). The x-axis represents institutional affiliations, displaying broad sectors (Public Sector vs. Think Tanks) on the left and a detailed breakdown of public sector entities on the right. The y-axis lists the first-level PACS code categories. The color scale indicates relative prominence: a value of 0.1 (white) represents the expected average distribution, while red color denotes higher-than-average citation intensity and blue color indicates lower-than-average citation intensity. Think Tanks exhibit a distinct preference for interdisciplinary areas of physics, whereas government documents show a relative under-representation of these interdisciplinary domains.}
    \label{fig:Institution}
\end{figure}

We first examine how the use of physics varies by institutional type. As shown in Figure \ref{fig:Institution}, Think Tanks exhibit the most pronounced preference for interdisciplinary areas of physics, dedicating 45.6\% of their citations to Category 0: General Physics and an additional 31.7\% to Category 8: Interdisciplinary Physics. These observations align with the role of the think tank in synthesizing broad strategic insights \cite{allern2020role}. In contrast, the public sector including government, Intergovernmental Organizations (IGOs), and legislative body displays specialized technical demands compared to think tanks. Government Agencies are significantly more likely to engage with materials science, with 34.4\% of citations concentrated in Condensed Matter Physics (Categories 6 and 7 combined), a pattern reflecting the agencies' focus on implementation-heavy technologies. IGOs show a distinct reliance on Category 2: Nuclear Physics (13.3\%), while Legislative bodies, though citing fewer physics papers, prioritize Category 4: Electromagnetism, Optics, Acoustics, Heat Transfer, Classical Mechanics, and Fluid Dynamics (13.7\%) (see~Table.S13--14). Additionally, within think tanks and IGOs, a small number of institutions produce a large share of documents in our dataset (see~Table.S11).

\begin{figure}[htbp]
    \centering
\includegraphics[width=0.8\linewidth]{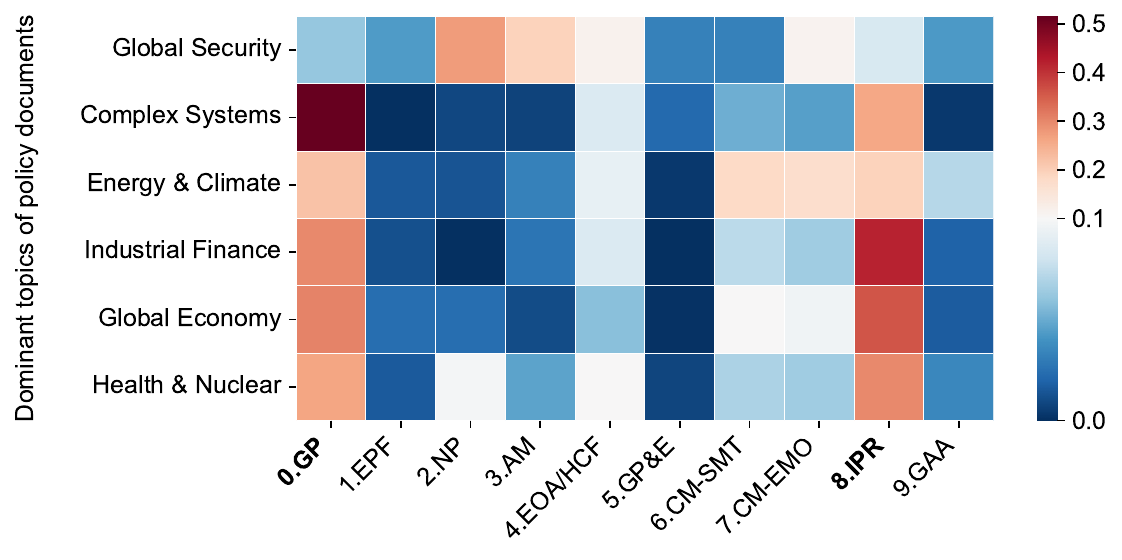}
    \caption{\textbf{Thematic preferences for physics subfield in policy documents.} The heatmap displays citation probabilities across six policy topics identified via LDA analysis: \textit{Global Security}, \textit{Complex Systems}, \textit{Energy and Climate}, \textit{Industrial Finance}, \textit{Global Economy}, and \textit{Health and Nuclear}. The x-axis lists first-level PACS code categories, while the y-axis represents the extracted topics. Each cell represents the probability that a given topic cites a specific first-level PACS category, where a value of 0.1 (white) corresponds to the expected baseline and red hues indicate higher citation intensity. Distinct disciplinary signatures emerge: \textit{Energy and Climate}, a domain heavily regulated by government agencies, draws primarily on Condensed Matter Physics. In contrast, \textit{Complex Systems}, a topic frequently associated with think tanks, relies predominantly on interdisciplinary areas of physics. These thematic distinctions mirror the institutional preference patterns observed in Fig.~\ref{fig:Institution}.}
\label{fig:heatmap}
\end{figure}

Institutional patterns are intrinsically linked to the specific policy topics addressed by different organizations. To map the relationship between topic and discipline, we apply Latent Dirichlet Allocation (LDA) to the policy corpus (See Materials and Methods), identifying six dominant themes: Global Security, Complex Systems, Energy and Climate, Industrial Finance, Global Economy, and Health and Nuclear (see~Table.S15). Figure \ref{fig:heatmap} visualizes the normalized distribution of citations across identified themes, revealing the context-dependent utility of physics knowledge. For instance, technical domains like \textit{Energy and Climate} prioritize condensed matter physics; papers classified under Category 6 (Structural) and Category 7 (Electronic) account for a combined 35.5\% of citations, underscoring the importance of materials science in energy production, storage, and climate dynamics. By contrast, integrative themes such as \textit{Health and Nuclear} (Category 0: 25.3\% and Category 8: 31.7\%) and \textit{Global Economy} (Category 0: 29.8\% and Category 8: 37.1\%) rely on broadly framed knowledge. Within these domains, interdisciplinary areas of physics account for the majority of citations, reflecting the interdisciplinary nature of these themes and their need for such integrative knowledge.

The observed thematic variation mirrors the institutional division of labor(see~Fig.S5). Institutions do not engage with all topics equally; rather, specific organizational mandates shape the type of physics consumed. Think tanks dominate the \textit{Complex Systems} theme, consistent with the sector's focus on analytically intensive, system-oriented research \cite{Furnas2025, teitz2009analysis}. Government agencies drive the discourse on \textit{Energy and Climate}, aligning with regulatory and implementation responsibilities \cite{yin2022public, an2022contingent}. Meanwhile, IGOs concentrate on \textit{Global Security}, reflecting the roles of institutions such as the United Nations Scientific Committee on the Effects of Atomic Radiation (UNSCEAR) and the International Atomic Energy Agency (IAEA) \cite{kjondal2021global}. Legislative bodies, focused on budgetary oversight, align closely with the economic implications of national security and defense policy \cite{natchez1973policy}. Thus, the mismatch discussed earlier is not a random artifact but a reflection of a functional ecosystem where specific institutions mobilize specific physics subfields to address distinct policy challenges.

\subsection{The Brokerage Role of Interdisciplinary Areas of Physics in Policy Domain}
While policy documents disproportionately cite interdisciplinary areas of physics, it remains to be seen how these fields are structurally integrated to support policy arguments. Do they function as isolated references or serve as the brokers linking distinct technical domains? In academic physics, interdisciplinary areas of physics have increasingly migrated toward the network core, functioning as glue both within physics and across its boundaries by linking distinct subfields and external domains, and exhibiting high influence across disciplinary contexts~\cite{Pan2012, sinatra2015century}. We extend this perspective to the policy domain to examine whether the logic of knowledge combination mirrors the structure of the broader scientific community or selectively amplifies the brokerage role of interdisciplinary physics.

To capture the structural relationships among physics subfields and how different areas of physics are bundled to address complex policy problems, we construct a weighted co-occurrence network using the first two digits of PACS codes. In this network, nodes represent distinct physics subfields; two nodes are connected if they are co-cited within a single policy report, with edge weights being the frequency of co-citations. This network can reveal how different areas of physics are bundled to address complex policy problems. For example, an IGO report on rural–urban links cites papers associated with PACS codes 87 (Biological and Medical Physics), 89 (Interdisciplinary Physics) and 05 (Statistical Physics), creating a web of connections that integrates biological, social and statistical frameworks within a single narrative. The full network consists of 68 nodes and 1,103 links. To distinguish systematic structural patterns from incidental noise, we extract the network backbone using the disparity filter (see Materials and Methods; $\alpha = 0.1$), yielding a core structure comprising 54 nodes and 135 links that preserves the network's essential topological features. We found that the nodes with the highest centrality included PACS 64, 89, 87, 02, and 05, with 89 in particular spanning broadly policy-relevant topics such as social organization, transportation, and environmental and energy studies (see~Table.S16).

\begin{figure}[h]
    \centering
\includegraphics[width=0.9\linewidth]{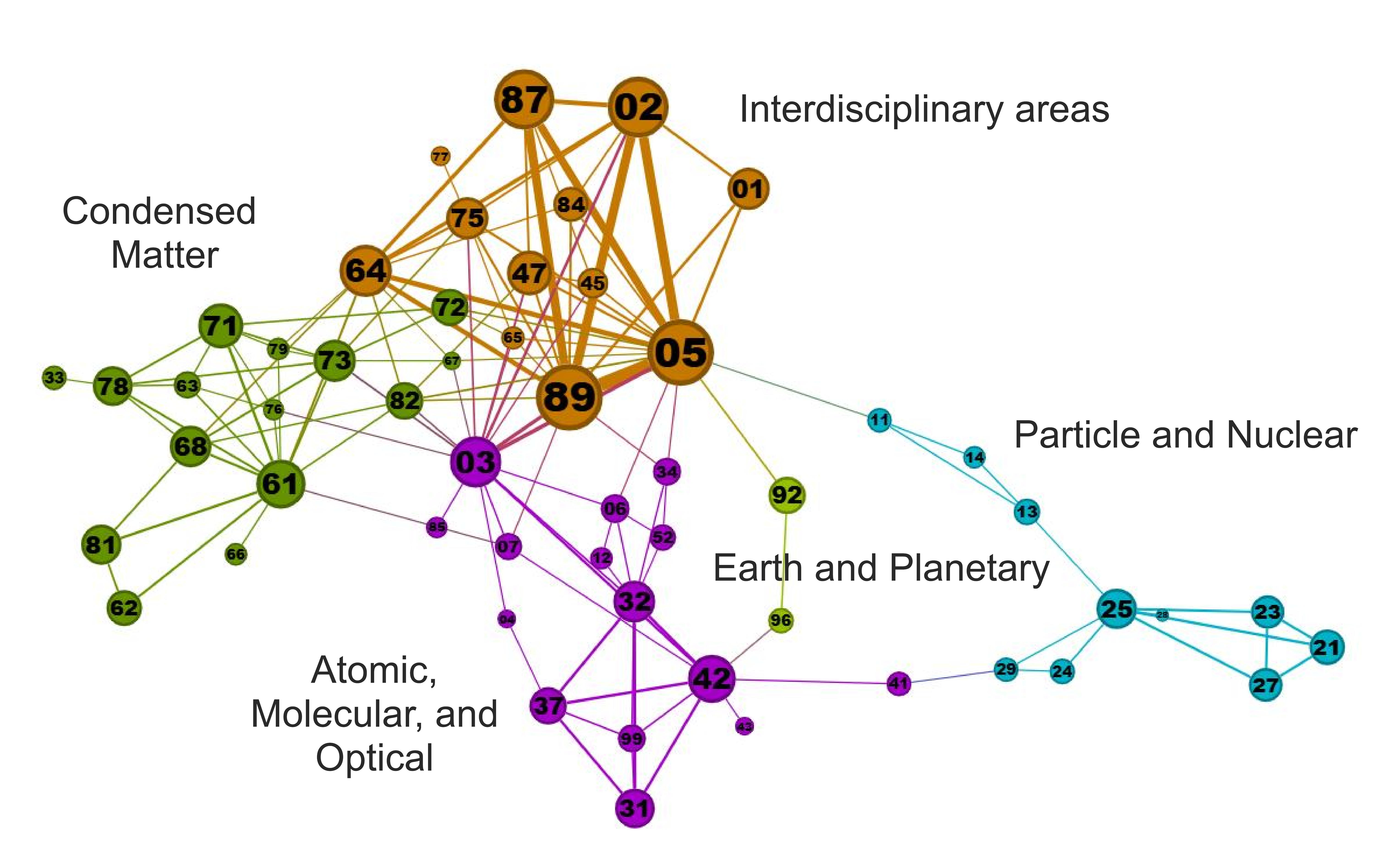}
    \caption{\textbf{Co-occurrence network of two-digit PACS codes in policy documents.} Nodes represent physics subfields, and links connect subfields cited together within the same policy document. Node size reflects the log-scaled frequency, while link thickness indicates co-occurrence intensity. To retain systematic structural patterns, the network is filtered using the disparity filter ($\alpha = 0.1$). We identify five distinct communities corresponding to major disciplinary clusters: \textit{Interdisciplinary Areas}, \textit{Condensed Matter}, \textit{Atomic, Molecular, \& Optical Physics}, \textit{Particle \& Nuclear Physics}, and \textit{Earth \& Planetary Sciences}. The topology reveals a hub-and-spoke structure where the \textit{Interdisciplinary areas of physics} communities (containing central nodes like 05 and 89) (see~Table.S17)form the integrative backbone, bridging the more specialized technical clusters.}
    \label{fig:network}
\end{figure}

We then apply community detection (see Materials and Methods) to the network, revealing a modular ``hub-and-community'' topology characterized by five distinct clusters. Interdisciplinary areas of physics occupy the structural core, acting as the primary connectors for specialized clusters such as Condensed Matter, Earth and Planetary Sciences, and Atomic, Molecular, and Optical Physics. These central hubs bridge otherwise disconnected domains, suggesting that broadly framed and interdisciplinary physics plays a critical organizing role in policy discourse. Centrality measures confirm this structural asymmetry: the interdisciplinary areas of the physics hub exhibit the highest average degree centrality ($0.14$) and the closeness centrality ($0.38$), indicating a strong brokerage function. By contrast, the Particle and Nuclear Physics cluster—comprising Elementary Particles (10) and Nuclear Physics (20)—shows substantially lower average degree ($0.06$) and betweenness centrality ($0.05$) (see~Table.S18). Despite strong internal cohesion, Particle and Nuclear Physics cluster remains relatively peripheral with limited external connectivity. The structural centrality of interdisciplinary fields is more pronounced in the policy domain. Compared with scientific co-occurrence networks (see~Fig.S7), policy networks place a sharper structural emphasis on interdisciplinary physics as integrative hubs.

Overall, interdisciplinary areas of physics occupy a central structural position within the policy network. By linking multiple communities, interdisciplinary areas of physics serve as a critical bridge among otherwise weakly connected research fields. A comparison with academic co-occurrence networks highlights a marked structural divergence. While interdisciplinary physics has become increasingly central within academic networks over time \cite{Pan2012, sinatra2015century}, policy-cited networks exhibit an even more selectively amplified brokerage role for interdisciplinary physics.

\vskip 2mm

\subsection{Visibility vs. Influence: The Role of Interdisciplinary Knowledge}

Building on the observation that policy documents selectively mobilize interdisciplinary physics, we examine whether these citation patterns translate into greater visibility for scientific research and subsequent policy impact. We conceptualize \textit{visibility} as the extent to which a scientific paper enters policy discourse, quantified by the number of citations the academic paper receives from policy documents. By contrast, we define the \textit{influence} of a policy document as its diffusion within the policy system, measured by the number of citations a policy document receives from subsequent policy documents. 

\begin{figure}[h]
    \centering
    \includegraphics[width=0.80\linewidth]{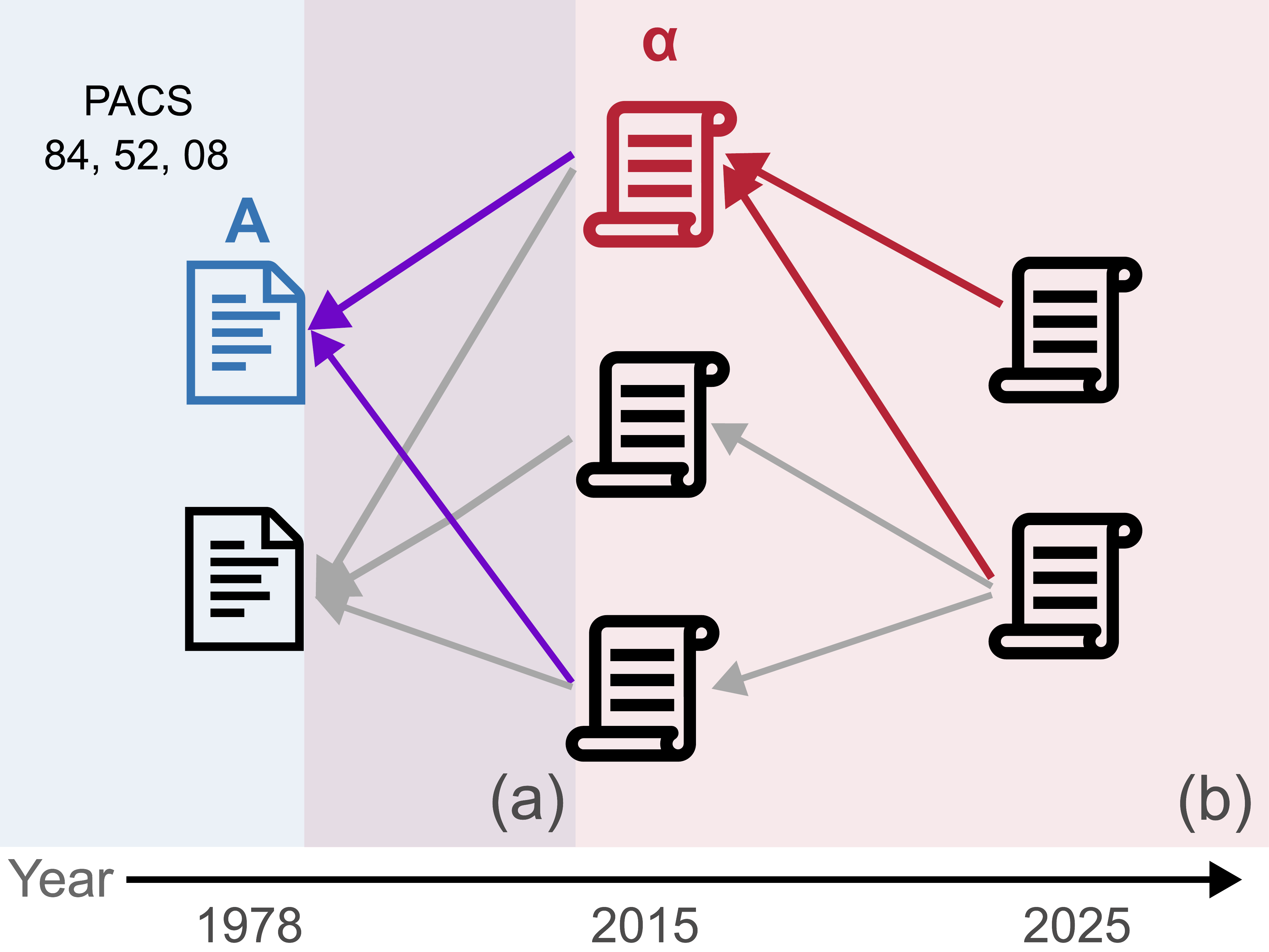}
    \caption{
      \textbf{ The schematic illustrates the distinction between \textit{visibility} and \textit{influence} at the science-policy interface. }The horizontal axis represents the temporal flow of citations (1978--2025). \textbf{(a)} \textit{Visibility} is measured by the uptake of science into policy (purple arrows); here, Paper A (associated with PACS categories 8, 5, and 0) is cited by two policy documents.\textbf{ (b)} \textit{Influence} is measured by policy-to-policy diffusion (red arrows), where policy document $\alpha$ accumulates citations from subsequent reports. Collectively, the diagram maps the translation of specific physics subfields into policy discourse and their subsequent propagation through the governance network.
    }
    \label{fig:Visibility_Influence}
\end{figure}

Figure~\ref{fig:Visibility_Influence} provides a schematic overview of our conceptual framework. In this example, the policy document $\alpha$ cites Academic Paper A, which is classified under PACS categories 8, 5 and 0. In Fig.~\ref{fig:Visibility_Influence} (a) illustrates \textit{visibility}, defined as the number of citations that paper A receives from policy documents. Panel (b) shows \textit{influence}, measured by the number of citations policy document $\alpha$ receives from subsequent policy reports. Crucially, through this citation linkage, document $\alpha$ inherits the disciplinary attributes of Paper A's discipline (categories 8, 5 and 0). To rigorously test these relationships, we estimate two Ordinary Least Squares (OLS) regression models. As citation distributions in both science and policy are characterized by heavy tails~\cite{vazquez2001statistics, wang2013quantifying} (see~Fig.S3), we employ log-transformed citation counts for all analyzes (see Materials and Methods for details).

First, we examine \textit{visibility} by analyzing the determinants of policy citation to individual scientific papers. In this analysis, the dependent variable captures the volume of policy attention that an academic paper receives, measured as the logarithm of the number of citations it accumulates from policy documents. The independent variables capture the core dimensions of scientific prominence and production, including the paper’s disruption score, log paper-to-paper citation counts, the presence of the top 1\% authors in the dataset, and team size. The regression includes fixed effects for the publication year and the journal to absorb temporal variation and venue-specific prestige(see Materials and Methods for details). 

\begin{figure}[h]
    \centering
\includegraphics[width=0.99\linewidth]{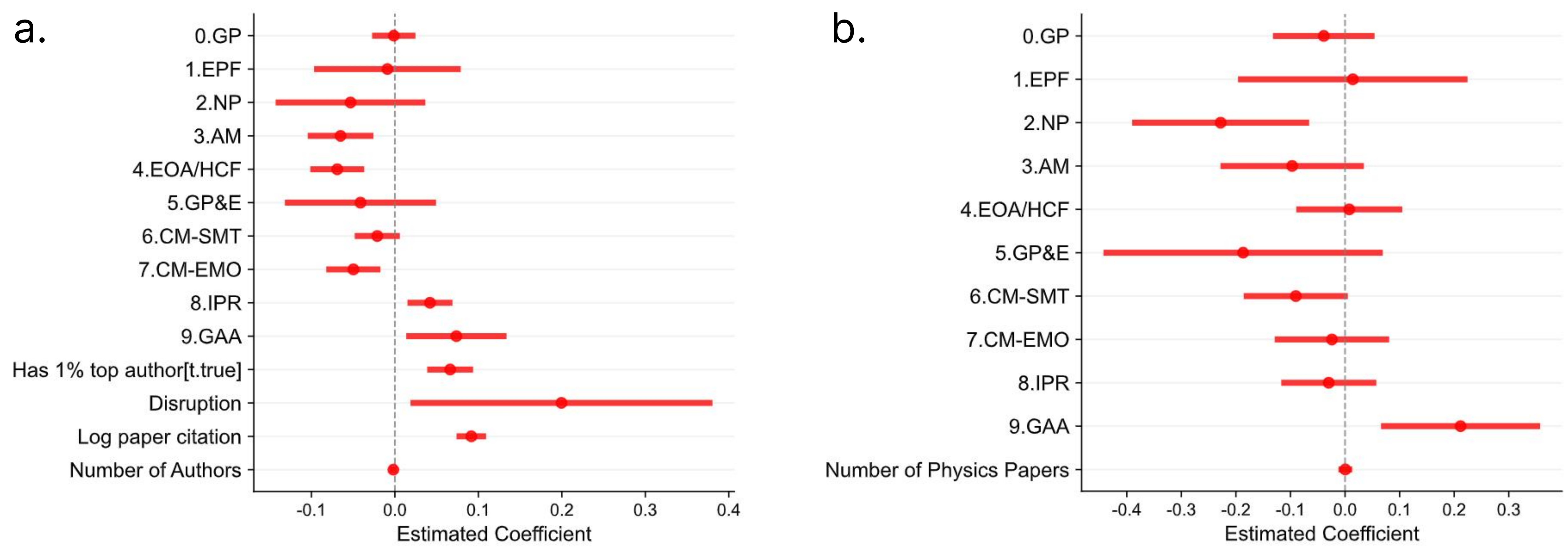}
    \caption{\textbf{Determinants of policy influence and scientific visibility.}
        \textbf{(a) Scientific visibility from policy:} OLS regression estimates predicting the visibility of an individual academic paper from policy documents. In this context, both Category 8 ($\beta = 0.04, p = 0.002$) and Category 9 ($\beta = 0.07, p = 0.016$) are positively associated with policy uptake. Scientific merit indicators are also strong predictors: disruption ($\beta = 0.20, p = 0.031$), academic citation count ($\beta = 0.09, p < 0.001$), and top-author presence ($\beta = 0.07, p < 0.001$) all significantly increase the likelihood of a paper being cited in policy.
        \textbf{(b) Policy influence:} OLS regression estimates predicting the citations a policy document receives from the wider policy ecosystem. Category 9: Geophysics, Astronomy, Astrophysics shows a significant positive association with downstream influence ($\beta = 0.212, p = 0.004$), whereas interdispinary area of physics are not: Category 8: Interdisciplinary Physics ($\beta = -0.03, p = 0.50$) and Category 0: General Physics ($\beta = -0.04, p = 0.41$).
        }
        \label{fig:regression}
\end{figure}

Figure~\ref{fig:regression}a illustrates the differential visibility of the physics subdisciplines. We find that Category 8 (Interdisciplinary Physics) and Category 9 (Geophysics/Astronomy) are associated with significantly higher rates of policy citation ($\beta = 0.04,\; p = 0.002$ and $\beta = 0.07,\; p = 0.016$, respectively). In contrast, specialized subfields appear to face a penalty: Category 3 ($\beta = -0.065,\; p = 0.001$), Category 4 ($\beta = -0.069,\; p < 0.001$), and Category 7 ($\beta = -0.050,\; p = 0.003$) all exhibit pronounced negative associations with policy uptake. Interestingly, Category 0: General Physics does not show a significant deviation ($\beta = -0.001,\; p = 0.919$) from the baseline. We also find that papers with higher scholarly citation counts are substantially more likely to be cited in policy texts ($\beta = 0.09,\; p < 0.001$). Similarly, disruptive papers ($\beta = 0.20,\; p = 0.031$) and those authored by elite scientists ($\beta = 0.07,\; p < 0.001$) exhibit a significantly higher likelihood of policy uptake, whereas team size shows no statistically significant effect ($\beta = -0.002,\; p = 0.371$) (see~Table.S1--2). Additionally, regression model including interaction among subfields models focusing on Category~8 and Category~9 provide little evidence of cross-subfield complementarities, showing limited evidence of cross-subfield complementarities (see~Table.S9--10).

Second, we examine the influence by analyzing the determinants of citations accruing to policy documents. In this analysis, the dependent variable captures the diffusion of a document within the policy system, measured as the natural logarithm of the number of citations it receives from subsequent policy documents. The independent variables capture the disciplinary composition and volume of the mobilized science, specifically through indicators for physics subfields and the count of cited papers. We further control for the aggregated scientific characteristics of the cited research—including the average disruption score, academic citations, and the presence of the top 1\% \ authors—as well as publication year, institutional type, and policy topic to absorb structural variations and temporal trends (see Materials and Methods for details).

Figure~\ref{fig:regression}b presents the estimated associations between cited physics subfields and subsequent policy impact. The results reveal substantial heterogeneity, indicating that the popularity of a mobilized subfield does not guarantee its downstream influence. Although Category 9 (Geophysics, Astronomy, and Astrophysics) is less frequently cited overall, policy documents referencing this field achieve significantly higher impact ($\beta = 0.212, p = 0.004$), corresponding to an approximate 24\% increase in downstream citations. In contrast, referencing Category 2 (Nuclear Physics) is associated with significantly diminished impact ($\beta = -0.23, p = 0.006$). Notably, despite serving as structural hubs within the co-occurrence network, interdisciplinary areas of physics do not show a statistically significant association with policy influence (Category 0: $\beta = -0.04, p = 0.41$ and Category 8:$\beta = -0.03, p = 0.50$, respectively). Moreover, alternative specifications including interaction terms reveal little evidence of cross-subfield complementarities (see~Table.S7--8). Furthermore, the sheer volume of physics knowledge cited is not a significant predictor ($\beta = 0.0005, p = 0.938$), suggesting that the policy impact is driven by the \textit{type} of knowledge mobilized rather than the \textit{intensity} of the citation.

The control variables further reveal that policy influence is stratified primarily by institutional authority and topic salience rather than scientific metrics. Institutional type is a powerful predictor: relative to the reference category, documents produced by Intergovernmental Organizations ($\beta = 0.663, p < 0.001$) and Think Tanks ($\beta = 0.328, p < 0.001$) garner substantially higher citation counts. Similarly, dominant policy topics strongly condition influence, with documents addressing \textit{Energy \& Climate} ($\beta = 0.495$) and the \textit{Global Economy} ($\beta = 0.445$) outperforming the baseline (see~Table.S6). In contrast, characteristics of the underlying scientific literature do not exhibit statistically significant associations with policy impact (see~Table.S3--5). This null result reinforces a fundamental disconnect: the metrics that signal value in the scientific community (novelty and academic prestige) do not seemingly drive the diffusion of knowledge within the policy sphere.

Synthesizing the results of the two regression models reveals a crucial distinction between \textit{entry} into the policy discourse (visibility) and downstream \textit{influence}. At the academic paper level, both scientific prominence and classification within Category 8: Interdisciplinary Physics facilitate entry into the policy corpus. However, this visibility does not automatically translate into systemic impact; while Category 8 is frequently mobilized, it does not drive subsequent policy-to-policy diffusion. In contrast, Category 9: Geophysics, Astronomy, and Astrophysics stands out as a unique case where high visibility ($\beta = 0.07$ in the paper-level model) translates directly into high downstream impact ($\beta = 0.21$ in the document-level model). Conversely, Category 2: Nuclear Physics exhibits a significant negative association with policy influence, while other subfields show no statistically significant effects. This suggests that while scientific merit opens the door, the specific disciplinary content determines whether that knowledge spreads through the policy network.

We find that the distinct influence of Category 9: Geophysics, Astronomy, and Astrophysics derives from the structural requirements of climate governance, a domain that demands cross-disciplinary and cross-border coordination~\cite{shaman2013fostering, bornmann2016policy}. Because effective climate policy relies on a unified evidence base, citations concentrate on a select group of synthesis reports—most notably the Assessments of the Intergovernmental Panel on Climate Change (IPCC). Within our dataset, such IPCC reports (and, to a lesser extent, outputs from high-profile think tanks such as the Heartland Institute) generate the majority of policy-to-policy citations (see~Table.S19--20). These synthesis documents function as standardization mechanisms~\cite{shaman2013fostering} by integrating Category 9 research into a consensus format repeatedly utilized across organizations and nations. Consequently, referencing Category 9 yields a higher downstream impact because the underlying physics becomes embedded within these high-circulation governance artifacts, generating a multiplier effect absent in more fragmented policy domains.

\section{Discussion}

Evidence-based policymaking functions not as a passive mirror of scientific production but as a highly selective process. We observe a fundamental mismatch between the "supply"of academic physics and its "demand" in policy. The divergence emerges because institutions mobilize scientific evidence strategically, selecting knowledge aligned with specific regulatory or strategic mandates rather than consuming physics based on academic volume \cite{Brian, Pawson}. Strategic filtering creates a distinct preference for integrative knowledge. Unlike scientific networks organized around deep specialization, the policy sphere favors broad frameworks capable of bridging disparate technical domains.

This preference for integrative knowledge manifests structurally in the co-citation network, where \textit{Interdisciplinary areas of physics} occupy central brokerage positions, linking otherwise isolated fields like Condensed Matter and Elementary Particles \cite{Cao, sinatra2015century}. Yet, our regression analysis reveals a paradox: structural centrality does not equate to systemic influence. While interdisciplinary knowledge acts as an essential ``internal glue" to constructing policy arguments, it fails to drive subsequent policy diffusion. Instead, high-impact transmission relies on specific domain-relevant evidence rather than structural brokering provided by interdisciplinary fields.

We observe the distinction between initial uptake and downstream influence most clearly in Category 9: Geophysics, which—despite lower initial visibility—commands high downstream influence. The divergence arises from the unique ecosystem of climate governance, where authoritative synthesis reports (e.g., IPCC assessments) standardize geophysics research into common reference points \cite{asatani2025influential, haunschild2016climate, bornmann2016policy}. Consequently, scientific utility follows a two-stage model: broad interdisciplinary framing facilitates \textit{entry} into policy discourse, while lasting \textit{influence} depends on integration into widely shared evidence infrastructures \cite{gu5293335universities, asatani2025influential, ba2023citation}.

Several limitations exist in our study on the scope and interpretability of these findings. First, our analysis is restricted by data availability and classification structures; we rely on the Overton database to capture explicit citations, and the aggregation of PACS codes—which are available only through 2016—constrains the temporal and conceptual resolution of the subfield analysis (see~Fig.S2). Second, relying on citations as a proxy for influence presents inherent validity challenges, as this metric captures formal acknowledgment while obscuring broader channels of impact, such as agenda setting, implementation, and the mediated role of advisory bodies \cite{Cao}. Just as academic citations are shaped by disciplinary norms and structural disparities rather than research quality alone \cite{Kim,radicchi2012reverse}, policy citations likely understate the full extent of scientific influence by failing to capture implicit forms of conceptual diffusion. Finally, our focus on physics necessarily limits generalizability, suggesting that future research should extend this framework to other domains to facilitate comparative analyzes. Despite these constraints, our study provides a systematic account of how scientific knowledge is selectively mobilized, elucidating the structural mismatch between scientific supply and policy demand while highlighting the critical role of interdisciplinary knowledge in the bridging of disconnected technical domains.

\bibliographystyle{unsrt}  
\bibliography{references}  

\clearpage

\section*{\centering\LARGE Supporting Information}
\addcontentsline{toc}{section}{Supporting Information}

\setcounter{table}{0}
\renewcommand{\thetable}{S\arabic{table}}

\setcounter{figure}{0}
\renewcommand{\thefigure}{S\arabic{figure}}

\vspace{5mm}

\subsection*{\centering\Large Regression Results}
Supplementary Tables 1--2 report \textit{Visibility} Ordinary Least Squares (OLS) regression results across alternative model specifications, while Tables 3--6 present the corresponding policy-level regressions. Across all specifications, the models are designed to identify associations between physics subfields and policy citation outcomes while accounting for institutional, topical, and scientific sources of variation.

\textit{Influence} is measured using policy-to-policy citation counts and analyzed on the logarithmic scale to accommodate their highly skewed distribution. Policy-level citations are characterized by substantial heterogeneity, with a small number of documents receiving a disproportionate share of citations. \textit{Visibility} regressions focus on the uptake of individual scientific papers in policy documents, measured by policy-to-paper citations, which are less dispersed but remain right-skewed.

Model specifications introduce control variables in conceptually structured blocks. Fixed effects for institutional type and policy topic capture systematic differences in how policy documents are produced and circulated. Additional controls for the volume and characteristics of cited research help distinguish subfield effects from broader patterns of scientific prominence and novelty. Across specifications, coefficient estimates for physics subfield indicators remain qualitatively stable.

Tables 7--10  further estimate interaction models to examine whether the associations between selected physics subfields and \textit{Influence} or \textit{Visibility} reflect cross-subfield complementarities. These interaction effects are generally modest and do not materially affect the main results. For this reason, the interaction specifications are included for completeness rather than for primary interpretation.

\vspace{5mm}
\begin{table}[htbp]
\centering
\caption{Visibility OLS Regression (Subfield Indicators Only)}
\label{tab:paper_level_ols_sig}
\begin{tabular}{lc}
\toprule
Variable & Main specification \\
\midrule
\\
Category 0
& 0.029** \\
& (2.27) \\

Category 2
& -0.062** \\
& (-2.38) \\

Category 3
& -0.051*** \\
& (-2.68) \\

Category 4
& -0.051*** \\
& (-3.02) \\

Category 6
& -0.032** \\
& (-2.25) \\

Category 7
& -0.048*** \\
& (-3.06) \\

Category 8
& 0.061*** \\
& (4.58) \\

Category 9
& 0.063** \\
& (2.12) \\
\midrule
\midrule
Year fixed effects & No \\
Other controls & No \\
Observations & 1,156 \\
R-squared & 0.087 \\
\bottomrule
\end{tabular}

\begin{flushleft}
\footnotesize
{Notes:} This table reports Visibility OLS regression results relating physics subfield indicators to paper visibility in policy documents, measured as log policy-to-paper citations. 
t statistics are reported in parentheses.
$^{***}p<0.01$, $^{**}p<0.05$, $^{*}p<0.1$.
\end{flushleft}
\end{table}

\begin{table}[htbp]
\centering
\caption{Visibility OLS Regression (Full Specification)}
\label{tab:category_policy}
\begin{tabular}{lc}
\hline
\\
Variable
 & Main specification \\
\hline
\\
Category 0
& -0.001 \\
& (-0.10) \\

Category 1
& -0.009 \\
& (-0.20) \\

Category 2
& -0.053 \\
& (-1.16) \\

Category 3
& -0.065*** \\
& (-3.25) \\

Category 4
& -0.069*** \\
& (-4.20) \\

Category 5
& -0.041 \\
& (-0.89) \\

Category 6
& -0.021 \\
& (-1.53) \\

Category 7
& -0.050*** \\
& (-3.01) \\

Category 8
& 0.042*** \\
& (3.06) \\

Category 9
& 0.074** \\
& (2.41) \\

Has top author
& 0.066*** \\
& (4.72) \\

Disruption
& 0.200** \\
& (2.17) \\

Log paper citation count
& 0.092*** \\
& (10.12) \\

Number of authors
& -0.002 \\
& (-0.89) \\
\midrule
\midrule
Year fixed effects & Yes \\
Other control & Yes \\
Observations & 1146 \\
R-squared & 0.235 \\
\hline
\end{tabular}

\begin{flushleft}
\footnotesize
{Notes:} This table presents the full Visibility regression specification, including physics subfield indicators, year fixed effects, journal fixed effects.
*** p $<$ 0.01, ** p $<$ 0.05, * p $<$ 0.1.
\end{flushleft}
\end{table}

\begin{table}[htbp]
\centering
\caption{Influence OLS Regression (Subfield Indicators Only)}
\label{tab:category_main_simplified}
\begin{tabular}{lc}
\toprule
\\
Variable & Main specification \\
\midrule
\\
Category 2
& -0.206** \\
& (-2.37) \\

Category 3
& -0.175** \\
& (-2.52) \\

Category 6
& -0.163*** \\
& (-3.18) \\

Category 9
& 0.272*** \\
& (3.63) \\
\midrule
\midrule
Year fixed effects & No \\
Other controls & No \\
Observations & 757 \\
R-squared & 0.063 \\
\bottomrule
\end{tabular}

\begin{flushleft}
\footnotesize
{Notes:} This table reports ordinary least squares regression results relating physics subfield indicators to policy influence, measured as log policy-to-policy citations. The specification includes only subfield indicators and serves as a reference point for subsequent models that introduce institutional, topical, and scientific controls.
t-statistics in parentheses.  
*** p $<$ 0.01, ** p $<$ 0.05, * p $<$ 0.1.
\end{flushleft}
\end{table}

\begin{table}[htbp]
\centering
\caption{Influence OLS Regression with Institution and Year Fixed Effects}
\label{tab:policy_ols}
\begin{tabular}{lc}
\toprule
\\
Variable & Main specification \\
\midrule
\\
Category 0
& 0.013 \\
& (0.29) \\

Category 1
& 0.094 \\
& (0.86) \\

Category 2
& -0.299*** \\
& (-3.62) \\

Category 3
& -0.204** \\
& (-3.11) \\

Category 4
& -0.004 \\
& (-0.09) \\

Category 5
& -0.158 \\
& (-1.18) \\

Category 6
& -0.064 \\
& (-1.30) \\

Category 7
& 0.013 \\
& (0.24) \\

Category 8
& -0.002 \\
& (-0.04) \\

Category 9
& 0.184** \\
& (2.60) \\

Number of physics papers
& -0.000 \\
& (-0.04) \\
\midrule
\midrule
Year fixed effects & Yes \\
Institution type fixed effects & Yes \\
Other controls & No \\
Observations & 752 \\
R-squared & 0.272 \\
\bottomrule
\end{tabular}

\begin{flushleft}
\footnotesize
{Notes:} This table reports Influence OLS regression results including year and institution type fixed effects. These fixed effects absorb systematic differences in policy production and citation practices across institutions and time, allowing subfield-level associations to be interpreted net of institutional and temporal heterogeneity.  
t-statistics in parentheses.  
*** p $<$ 0.01, ** p $<$ 0.05, * p $<$ 0.1.
\end{flushleft}

\end{table}

\begin{table}[htbp]
\centering
\caption{Influence OLS Regression (Full Specification)}
\label{tab:category_main}
\begin{tabular}{lc}
\hline
\\
Variable
 & Main specification \\
\hline
\\
Category 0
& -0.039 \\
& (-0.82) \\

Category 1
& 0.014 \\
& (0.13) \\

Category 2
& -0.228*** \\
& (-2.76) \\

Category 3
& -0.097 \\
& (-1.45) \\

Category 4
& 0.008 \\
& (0.16) \\

Category 5
& -0.187 \\
& (-1.43) \\

Category 6
& -0.090* \\
& (-1.86) \\

Category 7
& -0.024 \\
& (-0.45) \\

Category 8
& -0.030 \\
& (-0.67) \\

Category 9
& 0.212*** \\
& (2.85) \\

Number of physics papers
& 0.001 \\
& (0.08) \\
\midrule
\midrule
Year fixed effects & Yes \\
Other controls & Yes \\
Observations & 743 \\
R-squared & 0.330 \\
\hline
\end{tabular}

\begin{flushleft}
\footnotesize
{Notes:} This table presents the full Influence regression specification, including physics subfield indicators, year fixed effects, institution type fixed effects, policy topic fixed effects, and controls for the volume and characteristics of cited scientific research.
*** p $<$ 0.01, ** p $<$ 0.05, * p $<$ 0.1.
\end{flushleft}
\end{table}

\begin{table}[htbp]
\centering
\caption{Influence regression Control Variables}
\label{tab:controls_only}
\begin{tabular}{lc}
\hline
Variable & Controls only \\
\hline
\\
Policy source: IGO & 0.663*** \\
 & (12.82) \\
Policy source: Think Tank & 0.328*** \\
 & (5.99) \\
Policy source: Legislative Body & 0.170 \\
 & (0.92) \\
Dominant topic: Complex Systems & 0.265*** \\
 & (3.00) \\
Dominant topic: Energy \& Climate & 0.495*** \\
 & (6.09) \\
Dominant topic: Industrial Finance & 0.320*** \\
 & (3.63) \\
Dominant topic: Global Economy & 0.445*** \\
 & (5.75) \\
Dominant topic: Health \& Technology & 0.366*** \\
 & (4.54) \\
Has top author & -0.043 \\
 & (-0.83) \\
Average disruption & 0.089 \\
 & (0.29) \\
Average paper citations & -0.000 \\
 & (-0.34) \\
\hline
Year fixed effects & Yes \\
Category fixed effects & Yes \\
Observations & 743 \\
R-squared & 0.330 \\
\midrule
\midrule
\end{tabular}

\begin{flushleft}
\footnotesize
\textit{Notes:} This table reports coefficients from a specification including only control variables and fixed effects. 
t statistics are reported in parentheses.
$^{***}p<0.01$, $^{**}p<0.05$, $^{*}p<0.1$.
\end{flushleft}
\end{table}

\begin{table}[htbp]
\centering
\caption{Influence OLS Regression with Interaction Effects (Category 9)}
\label{tab:policy_interaction}
\begin{tabular}{lc}
\hline
 & Main specification \\
\hline
Category 0 & -0.029 \\
 & (-0.59) \\

Category 1 & 0.096 \\
 & (0.82) \\

Category 2 & -0.174** \\
 & (-2.00) \\

Category 3 & -0.080 \\
 & (-1.16) \\

Category 4 & 0.013 \\
 & (0.25) \\

Category 5 & -0.189 \\
 & (-1.44) \\

Category 6 & -0.092* \\
 & (-1.85) \\

Category 7 & -0.003 \\
 & (-0.06) \\

Category 8 & -0.006 \\
 & (-0.14) \\

Category 9 & 0.591*** \\
 & (3.65) \\

\hline
Category $\times$ Category 9 interactions &  \\
\hline

Category 1 $\times$ Category 9 & -0.538** \\
 & (-2.07) \\

Category 8 $\times$ Category 9 & -0.382** \\
 & (-1.99) \\

Other interactions & n.s. \\
\midrule
\midrule
Year fixed effects & Yes \\
Institution fixed effects & Yes \\
Topic fixed effects & Yes \\
Other controls & Yes \\
Observations & 743 \\
R-squared & 0.342 \\
\hline
\end{tabular}

\begin{flushleft}
\footnotesize
{Notes:} This table reports Influence regression results including interaction terms between Category 9 and selected physics subfields. Interaction terms are included to assess whether policy influence associated with Category 9 reflects cross-subfield complementarities.
t statistics are reported in parentheses.
$^{***}p<0.01$, $^{**}p<0.05$, $^{*}p<0.1$.
\end{flushleft}

\end{table}

\begin{table}[htbp]
\centering
\caption{Influence OLS Regression with Interaction Effects (Category 8)}
\label{tab:policy_interaction_cat8}
\begin{tabular}{lc}
\hline
 & Main specification \\
\hline
Category 0 & -0.077 \\
 & (-1.08) \\

Category 1 & -0.100 \\
 & (-0.79) \\

Category 2 & -0.289*** \\
 & (-3.03) \\

Category 3 & -0.190** \\
 & (-2.23) \\

Category 4 & -0.066 \\
 & (-0.91) \\

Category 5 & -0.198 \\
 & (-1.31) \\

Category 6 & -0.069 \\
 & (-0.88) \\

Category 7 & -0.080 \\
 & (-1.03) \\

Category 8 & -0.104 \\
 & (-1.17) \\

Category 9 & 0.289*** \\
 & (3.30) \\
\midrule
\midrule
Category $\times$ Category 8 interactions &  \\
\hline
Category 3 $\times$ Category 8 & 0.272* \\
 & (1.92) \\

Category 9 $\times$ Category 8 & -0.288* \\
 & (-1.76) \\

Other interactions & n.s. \\

\hline
Year fixed effects & Yes \\
Institution fixed effects & Yes \\
Topic fixed effects & Yes \\
Other controls & Yes \\
Observations & 743 \\
R-squared & 0.341 \\
\hline
\end{tabular}

\begin{flushleft}
\footnotesize
{Notes:}  This table reports policy-level regression results including interaction terms between Category 8 and selected physics subfields. The interaction specification examines whether interdisciplinary physics exhibits conditional associations with policy influence.
t statistics are reported in parentheses.
$^{***}p<0.01$, $^{**}p<0.05$, $^{*}p<0.1$.
\end{flushleft}
\end{table}

\begin{table}[htbp]
\centering
\caption{Visibility OLS Regression with Interaction Effects (Category 9)}
\label{tab:paper_interaction_cat9}
\begin{tabular}{lc}
\hline
 & Main specification \\
\hline
Category 0 & 0.008 \\
 & (0.59) \\
Category 1 & 0.023 \\
 & (0.48) \\
Category 2 & -0.030 \\
 & (-0.64) \\
Category 3 & -0.062*** \\
 & (-3.05) \\
Category 4 & -0.067*** \\
 & (-4.01) \\
Category 5 & -0.041 \\
 & (-0.86) \\
Category 6 & -0.017 \\
 & (-1.22) \\
Category 7 & -0.046*** \\
 & (-2.76) \\
Category 8 & 0.050*** \\
 & (3.62) \\
Category 9 & 0.341*** \\
 & (4.42) \\
\midrule
\midrule
Category $\times$ Category 9 interactions &  \\
\hline
Category 0 $\times$ Category 9 & -0.225*** \\
 & (-2.78) \\
Category 1 $\times$ Category 9 & -0.356*** \\
 & (-2.88) \\
Category 2 $\times$ Category 9 & -0.225** \\
 & (-2.09) \\
Category 6 $\times$ Category 9 & -0.250** \\
 & (-2.59) \\
Category 8 $\times$ Category 9 & -0.216** \\
 & (-2.45) \\
Other interactions & n.s. \\
\hline
Year fixed effects & Yes \\
Other control & Yes \\
Observations & 1,146 \\
R-squared & 0.251 \\
\hline
\end{tabular}

\begin{flushleft}
\footnotesize
{Notes:} This table reports Visibility OLS regression results with interaction terms between
Category 9 (Geophysics, Astronomy, and Astrophysics) and other physics subfields.
The dependent variable is the log number of policy-to-paper citations.
t statistics are reported in parentheses.
$^{***}p<0.01$, $^{**}p<0.05$, $^{*}p<0.1$.
\end{flushleft}
\end{table}

\begin{table}[htbp]
\centering
\caption{Visibility OLS Regression with Interaction Effects (Category 8)}
\label{tab:paper_interaction_cat8}
\begin{tabular}{lc}
\hline
 & Main specification \\
\hline
Category 0 & -0.022 \\
 & (-1.29) \\
Category 1 & -0.009 \\
 & (-0.19) \\
Category 2 & -0.051 \\
 & (-1.07) \\
Category 3 & -0.065*** \\
 & (-2.94) \\
Category 4 & -0.062*** \\
 & (-3.11) \\
Category 5 & -0.037 \\
 & (-0.78) \\
Category 6 & -0.015 \\
 & (-0.83) \\
Category 7 & -0.033* \\
 & (-1.68) \\
Category 8 & 0.039 \\
 & (1.42) \\
Category 9 & 0.110*** \\
 & (3.20) \\
\midrule
\midrule
Category $\times$ Category 8 interactions &  \\
\hline
Category 0 $\times$ Category 8 & 0.056** \\
 & (2.18) \\
Category 7 $\times$ Category 8 & -0.061* \\
 & (-1.82) \\
Category 9 $\times$ Category 8 & -0.187** \\
 & (-2.50) \\
Other interactions & n.s. \\
\hline
Year fixed effects & Yes \\
Other control & Yes \\
Observations & 1,146 \\
R-squared & 0.248 \\
\hline
\end{tabular}

\begin{flushleft}
\footnotesize
{Notes:} This table reports Visibility OLS regression results including interaction terms between Category 8 (Interdisciplinary Physics) and other physics subfields.
The dependent variable is the log number of policy-to-paper citations.
t statistics are reported in parentheses.
$^{***}p<0.01$, $^{**}p<0.05$, $^{*}p<0.1$.
\end{flushleft}
\end{table}

\clearpage
\subsection*{\centering\Large Authors \& Journals distribution}
$\text{HAS 1\% top author}_i$ is a simple indicator that captures whether a policy document refers to research produced by a very small group of highly productive scientists. Specifically, it equals one if the document cites at least one paper written by an author who belongs to the top 1\% of authors in our dataset in terms of the number of papers published, and zero otherwise. In our analytical sample, this top 1\% group consists of 128 authors out of 3,291, each appearing in at least 45 papers. Citations to such elite scientists are relatively rare. Approximately 23\% of policy documents cite at least one paper authored by a top 1\% scientist, while the majority do not. \\

By contrast, top authorship in policy documents is dominated by institutions rather than individuals. Among 714 distinct policy document authors, only nine belong to the top 1\% in terms of policy document production, and all nine are institutions rather than individual authors. These elite policy institutions include major intergovernmental organizations, government, and prominent Think tanks. When we include an indicator for elite policy institutions $\text{HAS 1\% top\_poli\_author}_i$, our results show that elite institutional status does not confer additional advantages for policy diffusion once broad institutional authority is taken into account. Conditional on institutional type, policy documents authored by top 1\% institutions are, on average, no more influential and in some specifications less influential than those produced by non-elite institutions within the same category.\\
\\
\begin{table}[htbp]
\centering
\caption{Top 1\% Policy Authors Ranked by the Number of Policy Documents}
\begin{tabular}{l r}
\hline
\textbf{Policy author} & \textbf{Policy documents} \\
\hline
Santa Fe Institute & 87 \\
United Nations & 62 \\
MITRE Corporation & 30 \\
OECD & 27 \\
NEA & 21 \\
State of California & 19 \\
Institute for New Economic Thinking & 9 \\
Institute for Defense Analysis & 8 \\
State of Idaho & 8 \\
\hline
\end{tabular}
\end{table}

\begin{table}[htbp]
\centering
\caption{Top 1\% Paper Authors Ranked by the Number of APS Papers Cited in Policy Documents}
\begin{tabular}{l r}
\hline
\textbf{Paper author} & \textbf{Policy-cited papers} \\
\hline
Carl E. Wieman & 48 \\
Mark Newman & 22 \\
Eric A. Cornell & 21 \\
Alex Zunger & 19 \\
Su-Huai Wei & 15 \\
H. Eugene Stanley & 14 \\
James P. Crutchfield & 12 \\
Charles B. Hanna & 11 \\
Vasiliki Plerou & 9 \\
J. M. Olson & 9 \\
David J. Wineland & 9 \\
Paolo Grigolini & 9 \\
Yong Zhang & 9 \\
Parameswaran Gopikrishnan & 9 \\
Deborah Jin & 8 \\
Shlomo Havlin & 8 \\
Luís A. Nunes Amaral & 7 \\
Michael R. Matthews & 7 \\
Mark A. Kasevich & 7 \\
Angelo Mascarenhas & 7 \\
Wayne M. Itano & 7 \\
Alex Punnoose & 7 \\
John F. Geisz & 7 \\
Lawrence W. Townsend & 7 \\
Helmut G. Katzgraber & 7 \\
Cristopher Moore & 7 \\
C. B. Hanna & 7 \\
J. L. Roberts & 7 \\
Shengbai Zhang & 7 \\
Luis Carlos Pardo & 6 \\
Neil R. Claussen & 6 \\
Lin-Wang Wang & 6 \\
Alexander L. Fetter & 6 \\
Jeffrey H. Shapiro & 6 \\
Seth Lloyd & 6 \\
Lenka Zdeborová & 6 \\
Brian Karrer & 6 \\
Dirk Helbing & 6 \\
Noah D. Finkelstein & 6 \\
Jeremy C. Smith & 6 \\
James C. Bergquist & 6 \\
Christopher Monroe & 6 \\
\hline
\end{tabular}
\end{table}

\begin{figure}[H]
  \centering
  \includegraphics[width=0.35\linewidth]{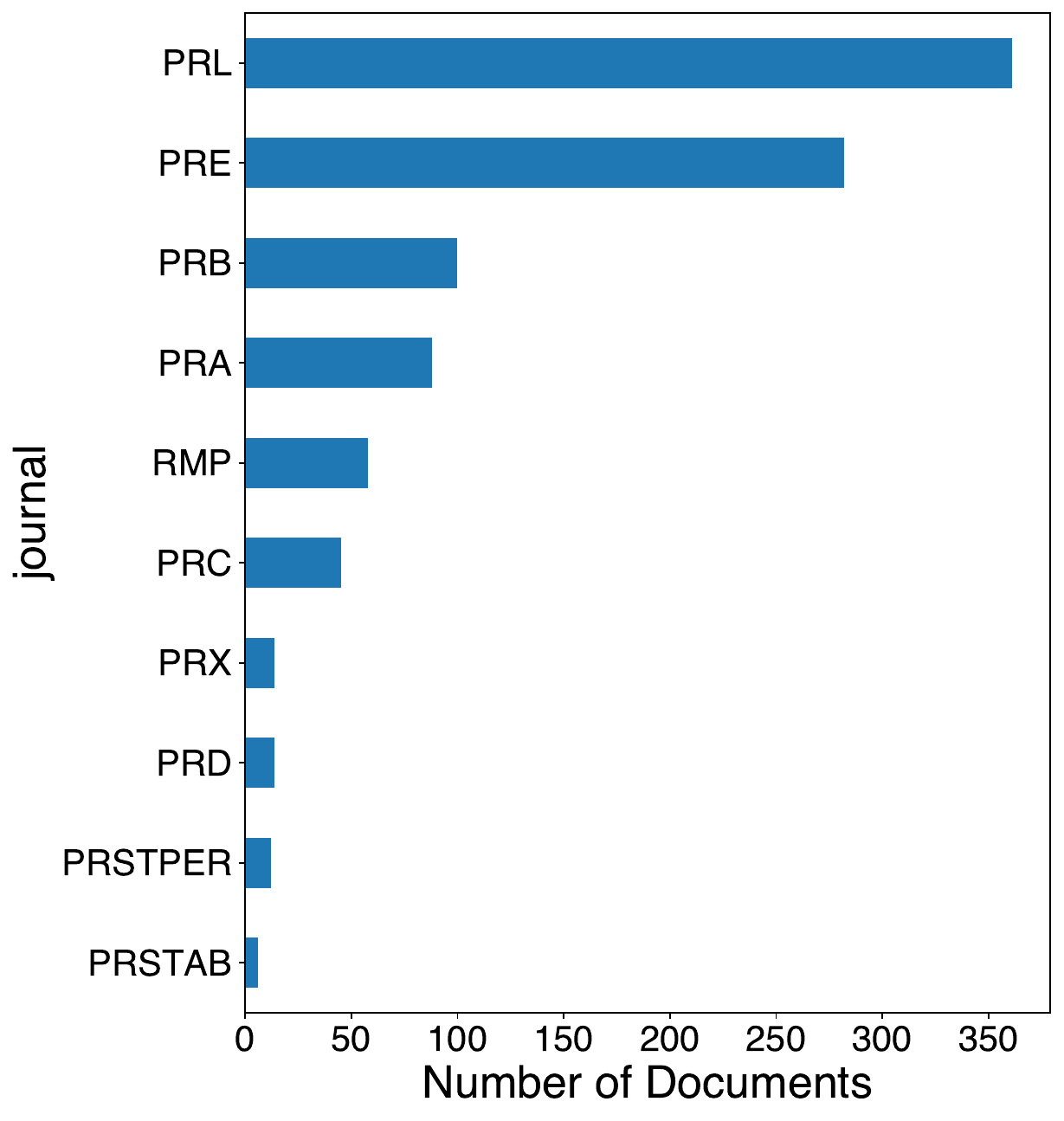}
  \caption{\textbf{Journal code distribution of APS papers cited in policy documents.} 
  Most cited papers are published in \emph{Physical Review Letters}, followed by \emph{Physical Review E} and \emph{Physical Review B}. 
  Journal codes are included as control variables in the paper-level regression analysis to account for systematic differences across journals.}
\end{figure}

\clearpage
\subsection*{\centering\Large Data coverage and citation distribution}
Tables 13–14 show that the total number of policy documents is broadly comparable across institutional types. Although government agencies account for the majority of documents in the full Overton database, only a small fraction of government produced documents cite at least one physics paper. In contrast, intergovernmental organizations produce a relatively higher share of physics citing policy documents, while the average number of cited physics papers per document is lower [1].\\

Figure 2 illustrates the asymmetric temporal coverage of the matched Overton–APS dataset. APS publications extend only through 2015, whereas Overton policy documents continue through 2025. Accordingly, the decline in policy documents citing physics after 2015, shown in panel (c), reflects data coverage limitations rather than a substantive decrease in policy engagement with physics research.\\

Figure 3 shows that citation links between policy documents and scientific papers are highly skewed, with a small number of documents and papers accounting for a disproportionate share of citations. Legislative bodies exhibit the highest average number of cited physics papers per policy document, but this pattern is driven by a very small number of cases and should be interpreted cautiously. \\
\\
\noindent\textbf{Reference}\\
\textbf{[1]} Szomszor M, Adie E. Overton: A bibliometric database of policy document citations. Quantitative science studies. 2022;3(3):624-50. \\
\\

\begin{table}[htbp]
\centering
\caption{Total number of policy documents by institution type.}
\label{tab:source_totals}
\begin{tabular}{lr}
\hline
\textbf{Institution Type} & \textbf{Total Documents} \\
\hline
Think Tank        & 228 \\
Government        & 252 \\
IGO               & 269 \\
Legislative Body  & 8   \\
\hline
\end{tabular}
\vskip 2mm
\begin{flushleft}
\footnotesize
\textit{Note:} The table reports the total number of policy documents by institution type. The full dataset contains 757 policy documents. All documents produced by think tanks, government agencies, and legislative bodies originate from the United States, whereas IGO documents are issued by international organizations.
\end{flushleft}
\end{table}

\begin{table}[htbp]
\centering
\caption{Average number of cited physics papers per policy document by institution type}
\label{tab:appendix_citation_intensity}
\begin{tabular}{l c}
\hline
Institution Type & Mean Physics Citations per Policy Document \\
\hline
Government & 2.51 \\
Intergovernmental Organization (IGO) & 1.68 \\
Legislative Body & 8.50 \\
Think Tank & 2.85 \\
\hline
\end{tabular}
\begin{flushleft}
\footnotesize
\textit{Note:} Values are computed as the total number of distinct APS physics papers cited by each institutional group divided by the number of unique policy documents produced by that group. The high average for legislative bodies reflects a small number of documents with intensive physics citation.
\end{flushleft}
\end{table}

\begin{figure}[H]
\centering 
\includegraphics[width=0.99\linewidth]{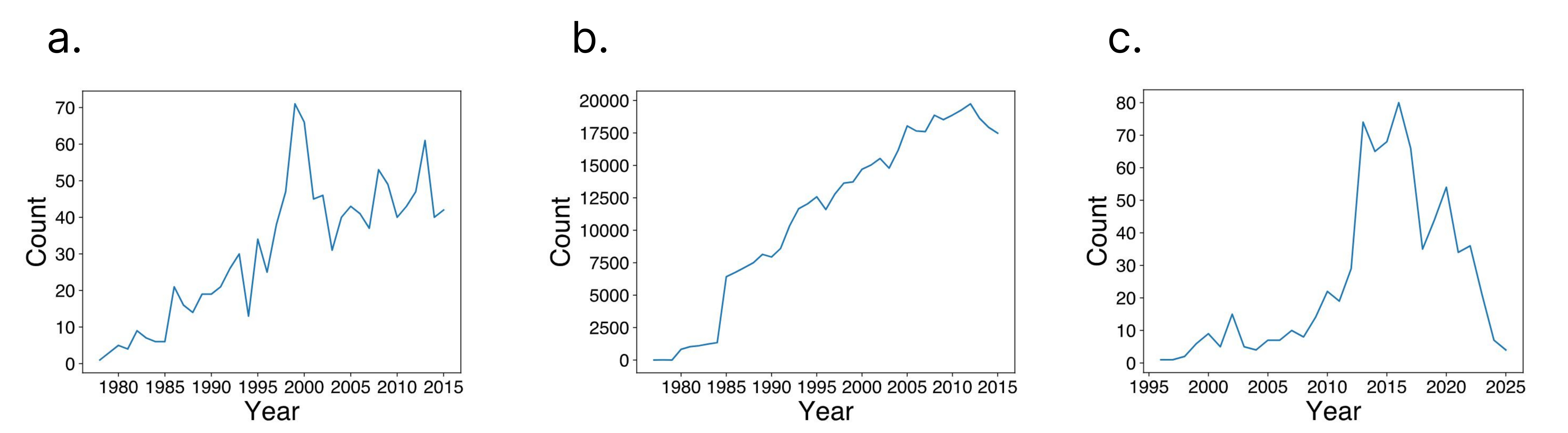} 
    \caption{\textbf{a.} Annual distribution of the full APS paper dataset, showing a marked increase in physics research output around the year 2000. \textbf{b.} Annual distribution of APS papers cited in policy documents, indicating a steady increase in the number of physics papers referenced in policy over time. \textbf{c.} Annual distribution of policy documents citing physics research. The apparent decline after 2015 reflects data coverage limitations, as the APS paper dataset includes publications only up to 2015.}
    \label{fig:Years}
\end{figure}

\begin{figure}[H]
\centering 
\includegraphics[width=0.99\linewidth]{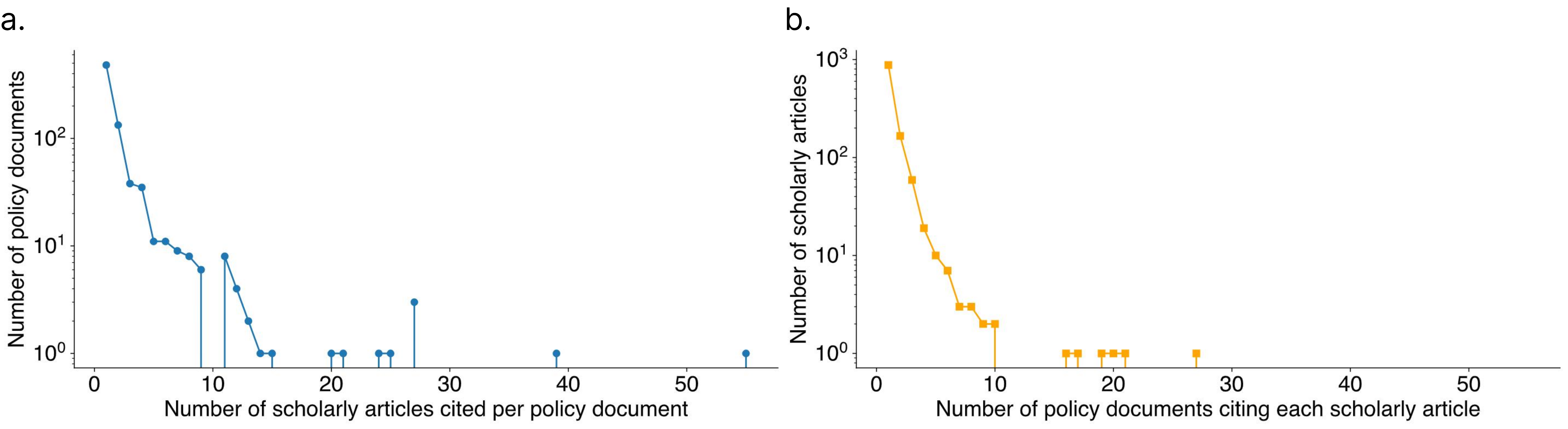} 
    \caption{\textbf{Distribution of citation links between policy documents and APS articles.} \textbf{a,} Number of APS articles cited per policy document. \textbf{b,} Number of policy documents citing each APS article. In the matched Overton–APS dataset, a single policy document cites up to 55 APS articles, while a single APS article is cited by as many as 27 policy documents, indicating that citation links are highly skewed.}
    \label{fig:Citation_dist}
\end{figure}

\clearpage
\subsection*{\centering\Large Topic modeling}

\begin{figure}[H]
\centering 
\includegraphics[width=0.99\linewidth]{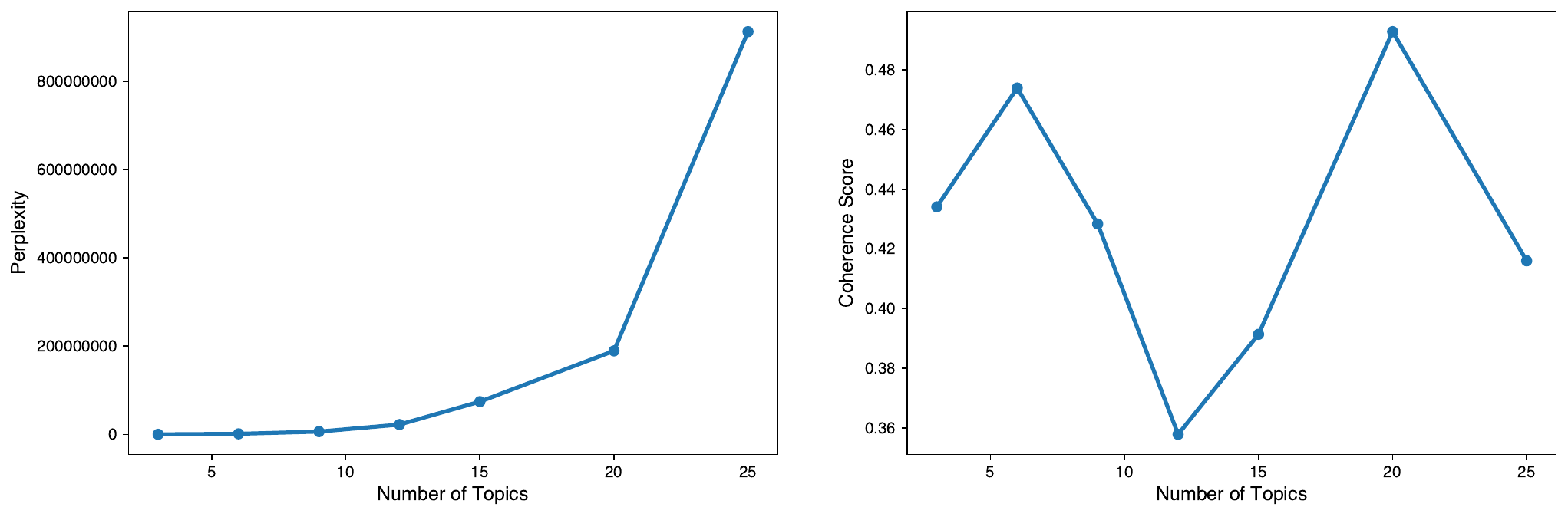} 
    \caption{\textbf{Selecting the number of topics.} Perplexity and coherence scores are shown for different numbers of topics. Lower perplexity indicates better statistical fit, whereas higher coherence reflects greater semantic interpretability. In this specification, perplexity increases monotonically with the number of topics, while topic coherence reaches a maximum at six topics, indicating that a six-topic model offers the most interpretable structure without additional gains in coherence.}
\label{fig:topic_coherence}
\end{figure}

\begin{table}[htbp]
\centering
\label{tab:topic_keywords}
\begin{tabular}{p{5cm} p{10cm}}
\hline
\textbf{Topic label} & \textbf{Representative keywords} \\
\hline
Global Security & council, security, resolution, United Nations, peace, Sudan, conflict, international, law, sanctions \\
\hline
Complex Systems & systems, processes, materials, data, theory, information, networks, sampling, modeling \\
\hline
Industrial Finance & financial, trade, industry, network, banks, countries, economic, growth, regional, integration \\
\hline
Energy \& Climate & climate, change, global emissions, energy, assessment, sustainability, mitigation, environment \\
\hline
Global Economy & economic, policy, global, trade, digital economy, socioeconomic, dynamics, supply, shocks, federal, policy \\
\hline
Health and Nuclear & Nuclear, medical, radiation, accelerator, pandemic, COVID, public health, fatty acids, nutrition, technology\\
\hline
\end{tabular}
\vspace{3mm}
\caption{Representative keywords for policy topics identified via topic modeling}
\end{table}

\begin{figure}[H]
\centering 
\includegraphics[width=0.99\linewidth]{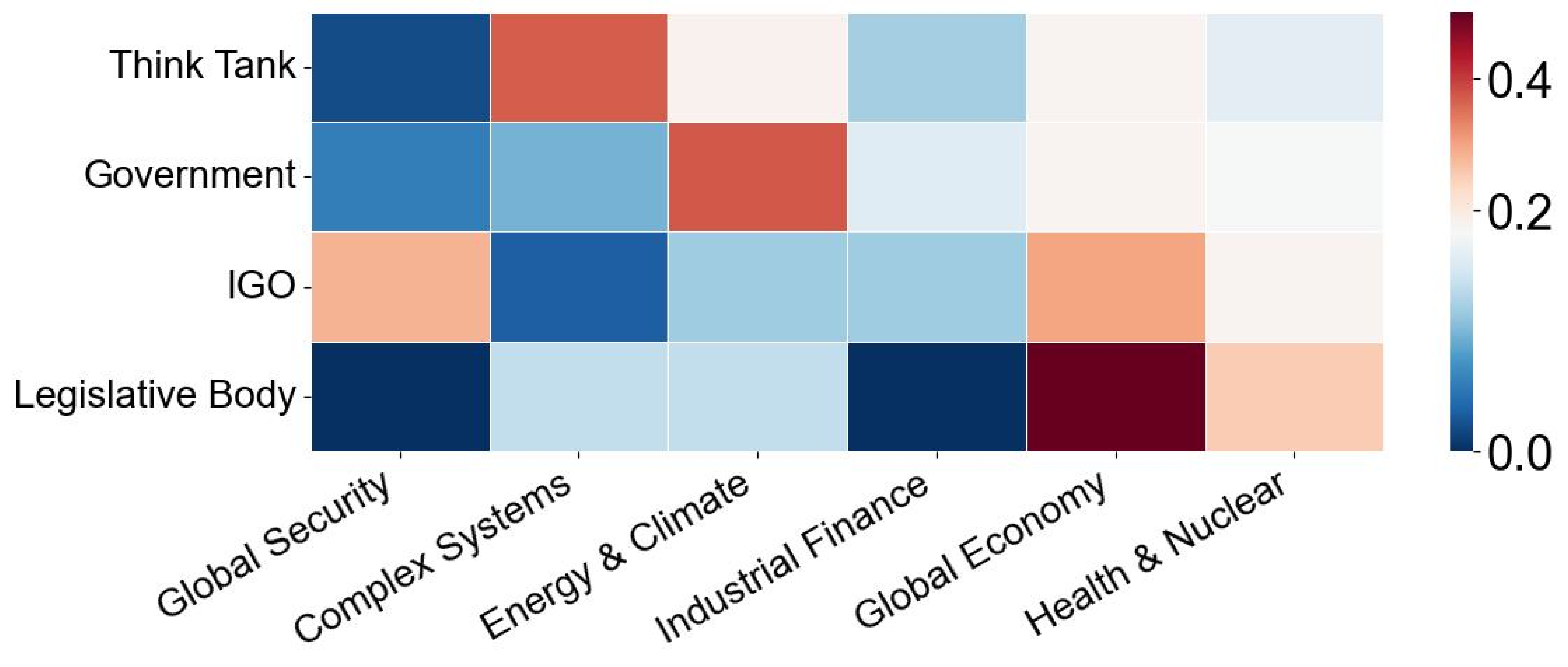} 
    \caption{\textbf{Institutional preferences for policy topics.} The heatmap displays the distribution of dominant topics across policy documents by institution type, with each column normalized to sum to one, thereby capturing the relative topic composition within each institution. Each cell represents the proportion of documents for which a given topic is the dominant theme. Government documents are most strongly concentrated in Energy and Climate, whereas think tanks exhibit a pronounced emphasis on Complex Systems. Intergovernmental organizations primarily focus on Global Security and the Global Economy. Legislative bodies show a higher concentration of documents related to the Global Economy and Health and Nuclear topics. These patterns align with the institution and topic specific PACS code preferences reported in Figures 3 and 4.}
    \label{fig:heatmap_topic_inst}
\end{figure}

\begin{figure}[H]
\centering 
\includegraphics[width=0.9\linewidth]{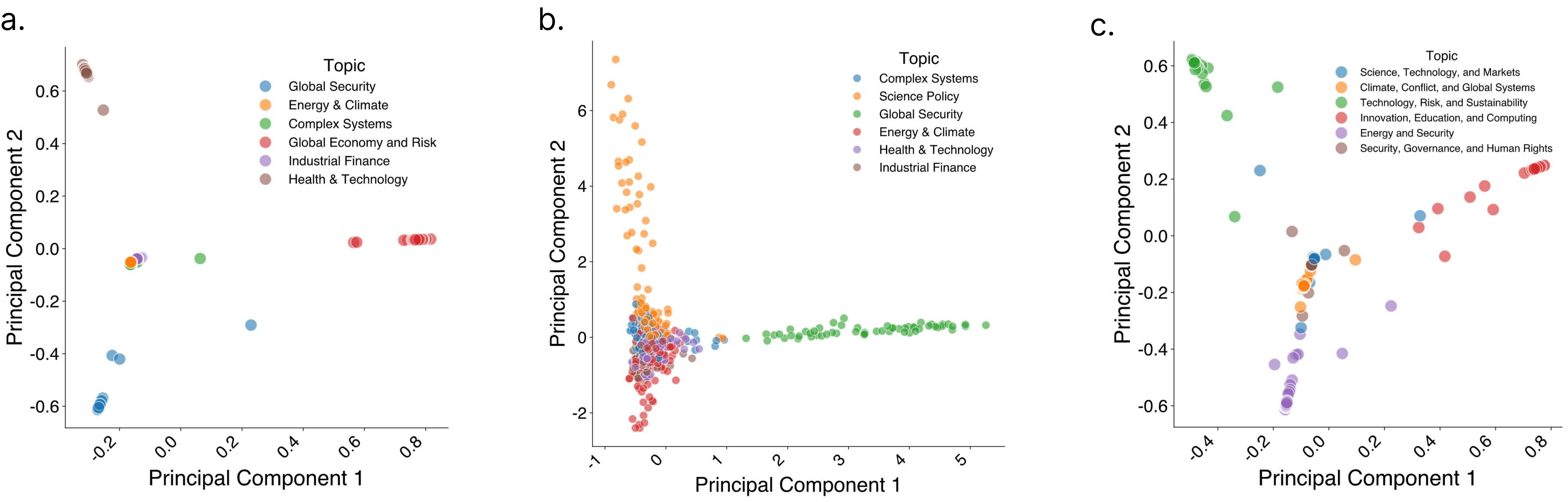} 
    \caption{PCA projection of document-level topic assignments using LDA and K-means. \textbf{a.}Latent Dirichlet Allocation (LDA) based on TF-IDF representations, where each document is projected according to its inferred topic probability distribution under different random initializations. \textbf{b.} K-means clustering, in which documents are assigned to clusters based on distance in the TF-IDF feature space. \textbf{c.} PCA projection of LDA using a CountVectorizer representation. Colors indicate dominant topic assignments. Overall, both methods produce comparable thematic groupings, while differences in cluster separation and topic labeling reflect the distinct assumptions underlying probabilistic topic modeling and distance-based clustering. The LDA model based on TF-IDF shows clearer topic separation compared to the CountVectorizer-based LDA.}
    \label{fig:LDA_분포}
\end{figure}

\clearpage
\subsection*{\centering\Large Network Analysis}
Building on the policy-document–level PACS two-digit co-occurrence network constructed in the main text, we further examine key nodes and community structures using centrality measures. For comparison, we additionally construct a PACS two-digit co-occurrence network at the level of individual scientific papers by using the full APS dataset. The nodes represent PACS codes that co-occur within the same article. The paper-level network comprises 83 nodes and 2,204 edges. Applying the same analytical pipeline as in the main text, including backbone extraction ($\alpha = 0.1$), Louvain community detection, and ForceAtlas2 layout, yields a backbone network with 70 nodes and 379 links. Comparing the two networks reveals both similarities and systematic differences. In both cases, interdisciplinary physics occupies a central hub position. However, in the paper-level network, the condensed matter community exhibits substantially higher centrality than in the policy-document network, and particle and nuclear physics are positioned closer to the network core.\\

\begin{table}[htbp]
\centering
\label{tab:community_pacs_degree}
\begin{tabular}{p{8cm} p{8cm}}
\hline
\textbf{Community} & \textbf{Constituent PACS codes} \\
\hline
Interdisciplinary areas of physics &
\textbf{64}, \textbf{89}, 87, \textbf{02}, \textbf{05},
\textbf{75}, 01, 45, 47, 77, 84 \\
\\
Condensed Matter &
\textbf{61}, 79, \textbf{73}, \textbf{71}, 81,
68, 62, 78, 63, 72, 82, 33, 76, 66 \\
\\
Atomic, Molecular, \& Optical Physics &
\textbf{03}, 41, \textbf{42}, 06, 32,
67, 34, 37, 99, 31, 07, 12, 43, 04, 85, 65, 52 \\
\\
Earth \& Planetary Sciences &
92, 96 \\
\\
Particle \& Nuclear Physics &
29, 13, 25, 14, 11,
24, 23, 21, 27, 28 \\
\hline
\end{tabular}
\label{fig:PACS_community}
\vspace{2mm}
\caption{\textbf{Community structure of two-digit PACS codes in the policy-level co-occurrence network.} Communities are identified using the Louvain algorithm. Boldface indicates the top ten PACS codes ranked by degree centrality in the full network.}
\end{table}

\begin{table}[htbp]
\centering
\begin{tabular}{c p{6cm} c c}
\hline
\textbf{PACS code} & \textbf{PACS category (PACS 2010)} & \textbf{Degree centrality} & \textbf{Betweenness centrality} \\
\hline
64 & Phase transitions and critical phenomena & 0.208 & 0.054 \\
89 & Other areas of applied and interdisciplinary physics & 0.264 & 0.064 \\
87 & Biological and medical physics & 0.113 & 0.000 \\
02 & Mathematical methods in physics & 0.151 & 0.001 \\
05 & Statistical physics, thermodynamics, and nonlinear dynamical systems & 0.340 & 0.254 \\
61 & Structure of solids and liquids; crystallography & 0.245 & 0.148 \\
75 & Magnetic properties and materials & 0.151 & 0.054 \\
03 & Quantum mechanics, field theories, and special relativity & 0.321 & 0.332 \\
41 & Electromagnetism; optics; acoustics; heat transfer; classical mechanics; fluid dynamics & 0.038 & 0.108 \\
29 & Experimental methods and instrumentation for elementary-particle and nuclear physics & 0.057 & 0.084 \\
42 & Optics & 0.170 & 0.207 \\
92 & Geophysics & 0.038 & 0.002 \\
96 & Solar system; planetary science & 0.038 & 0.008 \\
79 & Atomic and molecular collisions and interactions & 0.075 & 0.043 \\
73 & Electronic structure and electrical properties of surfaces, interfaces, thin films, and low-dimensional structures & 0.170 & 0.050 \\
71 & Electronic structure of bulk materials & 0.132 & 0.002 \\
81 & Materials science & 0.057 & 0.001 \\
68 & Surfaces and interfaces; thin films and nanosystems & 0.113 & 0.026 \\
62 & Mechanical and acoustical properties of condensed matter & 0.038 & 0.000 \\
78 & Optical properties, condensed-matter spectroscopy, and other interactions of radiation and particles with condensed matter & 0.113 & 0.039 \\
\hline
\end{tabular}
\caption{\textbf{Degree and betweenness centrality of top 20 PACS codes in the policy co-occurrence network}}
\label{fig:PACS_cent}
\end{table}

\begin{table}[htbp]
\centering
\label{tab:community_centrality_full}
\begin{tabular}{p{8cm} p{2cm} p{2cm} p{2cm}}
\hline
Community & Degree & Betweenness  & Closeness \\
\hline
Interdisciplinary Areas & 0.14 & 0.05 & 0.38 \\
Condensed Matter & 0.09 & 0.04 & 0.33 \\
Atomic, Molecular, \& Optical Physics & 0.09 & 0.07 & 0.35 \\
Particle \& Nuclear Physics & 0.06 & 0.05 & 0.25 \\
Earth \& Planetary Sciences & 0.04 & 0.00 & 0.33 \\
\hline
\end{tabular}
\vspace{2mm}
\caption{Community's centrality measures in the backbone network.}
\end{table}

\begin{figure}[H]
    \centering
    \includegraphics[width=0.7\linewidth]{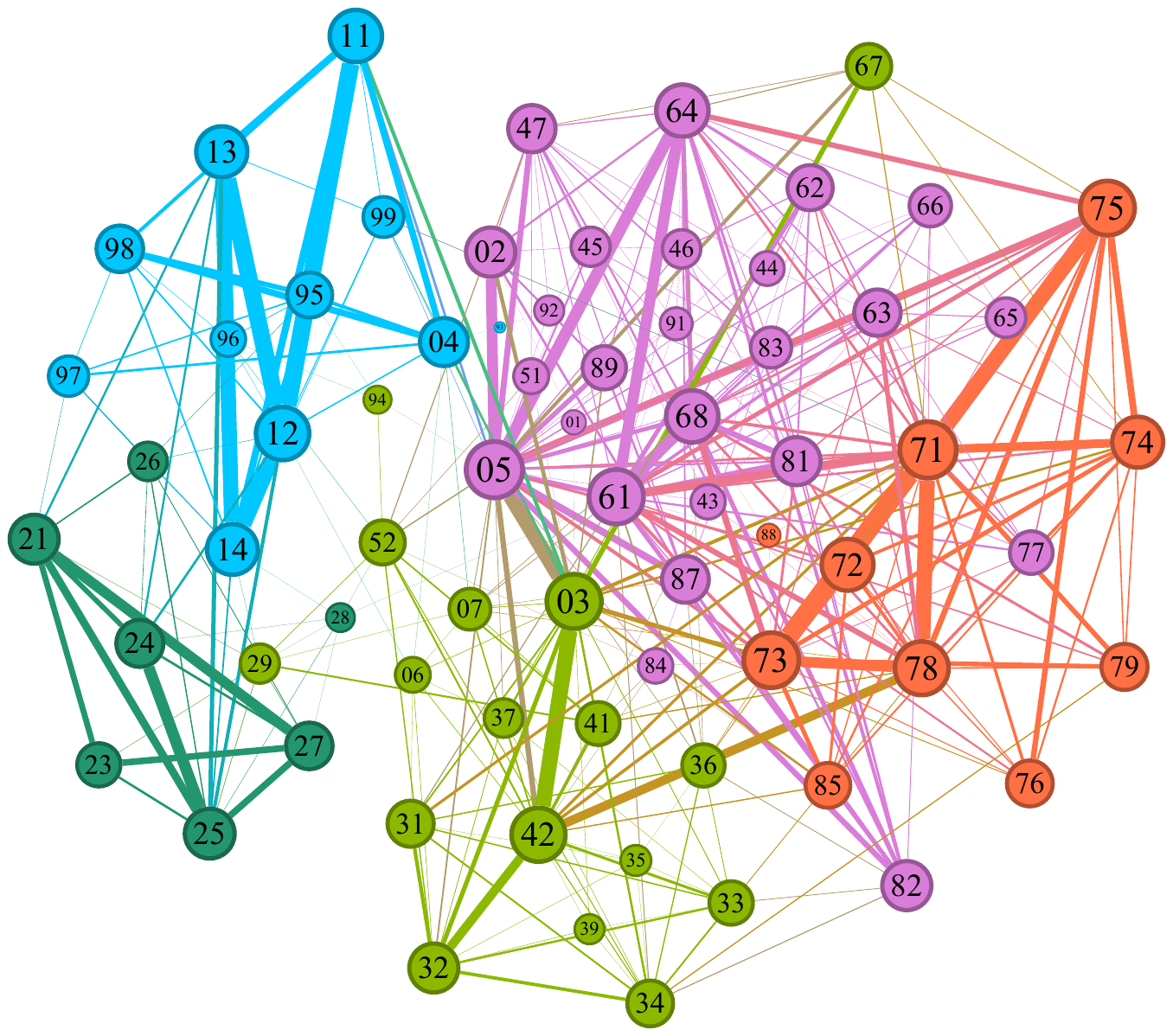}
    \caption{\textbf{Co-occurrence network of two-digit PACS codes in scientific papers.} Nodes represent physics subfields, and links connect subfields that co-occur within the same scientific paper. Node size reflects the log-scaled frequency, while link thickness indicates co-occurrence intensity. To retain systematic structural patterns, the network is filtered using the disparity filter ($\alpha = 0.1$). Community structure is identified using the Louvain algorithm and visualized with the ForceAtlas2 layout. Compared to the policy-document network, the paper-level topology exhibits stronger clustering within core disciplinary domains, with \textit{Condensed Matter Physics} and \textit{Particle} \& \textit{Nuclear Physics} occupying more central positions, while \textit{Interdisciplinary areas of physics} continue to function as an integrative hub.}
    \label{fig:network_paper}
\end{figure}

\newpage
\begin{sidewaystable}[p]
\centering
\caption{Highly cited policy documents by category}
\small
\begin{tabular}{c p{7cm} c c c}
\hline
\textbf{Category} & \textbf{Policy document title} & \textbf{Institution} & \textbf{Year} & \textbf{Citation} \\
\hline
\hline
\multirow{3}{*}{0} 
& AR5 Climate Change 2013: The Physical Science Basis & IPCC & 2013 & 5905 \\
& AR4 Climate Change 2007: The Physical Science Basis & IPCC & 2007 & 5043 \\
& AR6 Climate Change 2021: The Physical Science Basis & IPCC & 2021 & 4273 \\
\hline
\multirow{2}{*}{1}
& Fats and Fatty Acid in Human Nutrition & FAO & 2010 & 125 \\
& Grasas y Ácidos Grasos en Nutrición Humana & FAO & 2010 & 15 \\
\hline
\multirow{2}{*}{2}
& The Future of Nuclear Power in China & Carnegie Endowment for International Peace & 2018 & 16 \\
& Handbook on Lead-bismuth Eutectic Alloy and Lead Properties & OECD NEA & 2015 & 8 \\
\hline
\multirow{2}{*}{3}
& Quantum Computing and Communications: Status and Prospects & U.S. GAO & 2021 & 26 \\
& Commercial and Military Applications of Quantum Technology & RAND Corporation & 2021 & 22 \\
\hline
\multirow{3}{*}{4}
& AR6 Climate Change 2021: The Physical Science Basis & IPCC & 2021 & 4273 \\
& TAR Climate Change 2001: The Scientific Basis & IPCC & 2001 & 1559 \\
& The Assessment Report on Pollinators, Pollination and Food Production & IPBES & 2017 & 189 \\
\hline
\multirow{2}{*}{5}
& PENELOPE 2014: A Code System for Monte Carlo Simulation of Electron and Photon Transport & OECD NEA & 2015 & 3 \\
& PENELOPE 2011: A Code System for Monte Carlo Simulation of Electron and Photon Transport & OECD NEA & 2012 & 1 \\
\hline
\multirow{3}{*}{6}
& Global Resources Outlook 2019: Natural Resources for the Future We Want & UNEP & 2019 & 170 \\
& Future Global Shocks & OECD & 2011 & 82 \\
& Social Interactions in Pandemics: Fear, Altruism, and Reciprocity & NBER & 2020 & 21 \\
\hline
\multirow{3}{*}{7}
& Renewable Energy Sources and Climate Change Mitigation & IPCC & 2011 & 967 \\
& OECD Digital Economy Outlook 2020 & OECD & 2020 & 256 \\
& Global Resources Outlook 2019: Natural Resources for the Future We Want & UNEP & 2019 & 170 \\
\hline
\multirow{3}{*}{8}
& AR5 Climate Change 2014: Mitigation of Climate Change & IPCC & 2014 & 3161 \\
& Managing the Risks of Extreme Events and Disasters to Advance Climate Change Adaptation & IPCC & 2012 & 2911 \\
& The Assessment Report on Pollinators, Pollination and Food Production & IPBES & 2017 & 189 \\
\hline
\multirow{6}{*}{9}
& AR5 Climate Change 2013: The Physical Science Basis & IPCC & 2013 & 5905 \\
& AR4 Climate Change 2007: The Physical Science Basis & IPCC & 2007 & 5043 \\
& AR6 Climate Change 2021: The Physical Science Basis & IPCC & 2021 & 4273 \\
& Managing the Risks of Extreme Events and Disasters to Advance Climate Change Adaptation & IPCC & 2011 & 2911 \\
& TAR Climate Change 2001: The Scientific Basis & IPCC & 2001 & 1559 \\
& Climate Change Reconsidered II: Physical Science & Heartland Institute & 2009 & 32 \\
\hline
\end{tabular}
\end{sidewaystable}

\begin{sidewaystable}[htbp]
\centering
\small
\caption{Physics papers cited by highly cited policy documents}
\begin{tabular}{c p{15cm} l}
\hline
\textbf{Category} & \textbf{Article title} & \textbf{DOI} \\
\hline
\hline
\multirow{3}{*}{0}
& From Granger causality to long-term causality: Application to climatic data & 10.1103/physreve.80.016208 \\
& Low Cloud Properties Influenced by Cosmic Rays & 10.1103/physrevlett.85.5004 \\
& Influence of Cosmic Rays on Earth's Climate & 10.1103/physrevlett.81.5027 \\
\hline
\multirow{3}{*}{1}
& Leptonic CP violation in a two parameter model & 10.1103/physrevd.71.093005 \\
& Why the universe is just so & 10.1103/revmodphys.72.1149 \\
& CODATA recommended values of the fundamental physical constants: 2006 & 10.1103/revmodphys.80.633 \\
\hline
\multirow{2}{*}{2}
& Physics design of an accelerator for an accelerator-driven subcritical system & 10.1103/physrevstab.16.080101 \\
& New potentialities of the Liège intranuclear cascade model for reactions induced by nucleons and light charged particles & 10.1103/physrevc.87.014606 \\
\hline
\multirow{2}{*}{3}
& 14-Qubit Entanglement: Creation and Coherence & 10.1103/physrevlett.106.130506 \\
& Gate-count estimates for performing quantum chemistry on small quantum computers & 10.1103/physreva.90.022305 \\
\hline
\multirow{2}{*}{4}
& Insect flight dynamics: Stability and control & 10.1103/revmodphys.86.615 \\
& Characterizing the development of sectoral gross domestic product composition & 10.1103/physreve.88.012804 \\
\hline
\multirow{2}{*}{5}
& Ratio of positron to electron bremsstrahlung energy loss: An approximate scaling law & 10.1103/physreva.33.3002 \\
& Generating multi-GeV electron bunches using single stage laser wakefield acceleration in a 3D nonlinear regime & 10.1103/physrevstab.10.061301 \\
\hline
\multirow{3}{*}{6}
& First-principles study of the polar (111) surface of Fe$_3$O$_4$ & 10.1103/physrevb.74.035409 \\
& Breakdown of the Internet under Intentional Attack & 10.1103/physrevlett.86.3682 \\
& Epidemic dynamics and endemic states in complex networks & 10.1103/physreve.63.066117 \\
\hline
\multirow{2}{*}{7}
& High Efficiency Carrier Multiplication in PbSe Nanocrystals: Implications for Solar Energy Conversion & 10.1103/physrevlett.92.186601 \\
& Experimental Realization of a Quantum Support Vector Machine & 10.1103/physrevlett.114.140504 \\
\hline
\multirow{3}{*}{8}
& Superlinear scaling for innovation in cities & 10.1103/physreve.79.016115 \\
& A Complexity View of Rainfall & 10.1103/physrevlett.88.018701 \\
& Insect flight dynamics: Stability and control & 10.1103/revmodphys.86.615 \\
\hline
\multirow{3}{*}{9}
& From Granger causality to long-term causality: Application to climatic data & 10.1103/physreve.80.016208 \\
& Low Cloud Properties Influenced by Cosmic Rays & 10.1103/physrevlett.85.5004 \\
& Influence of Cosmic Rays on Earth's Climate & 10.1103/physrevlett.81.5027 \\
\hline
\end{tabular}
\end{sidewaystable}

\end{document}